\documentclass[aps,prb,showpacs,nofootinbib,twocolumn,amsmath,amssymb,superscriptaddress]{revtex4-1}

\usepackage{latexsym}
\usepackage{graphicx}
\usepackage{times,psfrag,subfigure}
\usepackage{enumerate}
\usepackage{amsmath}
\usepackage{dsfont}
\usepackage{dcolumn}
\usepackage{bm,bbm}
\usepackage{color}
\usepackage{latexsym,amsmath,amssymb,bm,euscript}
\usepackage{dsfont}
\usepackage{textcomp}
\usepackage{tabularx}
\usepackage{setspace}
\usepackage{ctable}
\usepackage{sidecap}
\usepackage{placeins}
\usepackage{threeparttable}
\usepackage{multirow}
\usepackage[hidelinks]{hyperref}

\let \oldbm \bm
\renewcommand{\vec}[1]{\oldbm{#1}}

\def\bk{{\vec k}}
\def\td{q}
\def\be{{\vec e}}

\def\bn{{\vec n}}
\def\bm{{\vec m}}

\def\br{{\vec r}}
\def\bgamma{{\boldsymbol \gamma}}
\def\bnabla{{\boldsymbol \nabla}}
\def\balpha{{\boldsymbol \alpha}}

\def\tr{\mathop{\mathrm{tr}}}
\def\sgn{\mathop{\mathrm{sgn}}}
\def\Z{\mathds{Z}}
\def\T{\mathcal{T}}
\def\C{\mathcal{C}}
\def\P{\mathcal{P}}
\def\S{\mathcal{S}}
\def\I{\mathcal{I}}

\def\H{\mathcal{H}}
\def\K{\mathcal{K}}
\def\M{\mathcal{M}}

\def\A{\mathcal{A}}

\newcommand{\beq}{\begin{equation}}
\newcommand{\eeq}{\end{equation}}
\newcommand{\beqarray}{\begin{eqnarray}}
\newcommand{\eeqarray}{\end{eqnarray}}

\newcommand{\bsigma}{\mbox{\boldmath$\sigma$}}

\allowdisplaybreaks

\begin{document}

\title{Higher-order topological insulators and superconductors protected by inversion symmetry}

\date{\today}

\begin{abstract}
We study surface states of topological crystalline insulators and superconductors protected by inversion symmetry. These fall into the category of ``higher-order'' topological insulators and superconductors which possess surface states that propagate along one-dimensional curves (hinges) or are localized at some points (corners) on the surface. We provide a complete classification of inversion-protected higher-order topological insulators and superconductors in any spatial dimension for the ten symmetry classes by means of a layer construction. We discuss possible physical realizations of such states starting with a time-reversal invariant topological insulator (class AII) in three dimensions or a time-reversal invariant topological superconductor (class DIII) in two or three dimensions. The former exhibits one-dimensional chiral or helical modes propagating along opposite edges, whereas the latter hosts Majorana zero modes localized to two opposite corners. Being protected by inversion, such states are not pinned to a specific pair of edges or corners thus offering the possibility of controlling their location by applying inversion-symmetric perturbations such as magnetic field.
\end{abstract}

\author{Eslam Khalaf}
\affiliation{Max Planck Institute for Solid State Research, Heisenbergstr.\ 1, 70569 Stuttgart, Germany}
\affiliation{Department of Physics, Harvard University, Cambridge MA 02138}

\maketitle

\section{Introduction} 
A topological insulator (TI) or superconductor (TSC) is characterized by a gapped bulk spectrum with gapless states on any given surface \cite{Molenkamp, Hasan10, Moore09, Qi11}. The stability of these gapless surface states in a TI/TSC usually relies on the presence of local symmetries which include time-reversal symmetry (TRS) $\T$, particle-hole symmetry (PHS) $\P$, and their combination $\S$ (usually called chiral symmetry). According to the presence or absence of these symmetries, gapped Hamiltonians have been classified  into ten different symmetry classes by Altland and Zirnbauer (AZ) \cite{Altland97}. In any given dimension, five of these ten classes correspond to a TI/TSC \cite{Kitaev09, Schnyder09, Ryu10}. 

The concept of a TI/TSC can be extended to include gapped phases protected by crystalline symmetries which map different points in space to each other such as mirror, rotation, or inversion symmetries. These phases are called topological crystalline insulators (TCIs) \cite{Fu11, Hsieh12} and, unlike conventional TIs/TSCs, do not necessarily host gapless surface states on any given surface. Instead, surface states are only expected on surface planes which preserve the crystalline symmetry.

Recently, several works considered a class of TCIs which instead host surface states localized to the edges (hinges) or corners of a physical sample \cite{Schindler17, Langbehn17, Benalcazar16, Song17, Fang17, Khalaf17}; these were dubbed higher-order TIs/TSCs \cite{Schindler17, Langbehn17}. The surface of a $k$-th order TI/TSC in $d$ dimensions is gapped except for a $(d-k)$-dimensional region which hosts gapless modes. For example, a second-order TI in three dimensions hosts one-dimensional (1D) propagating modes localized to the sample hinges on an otherwise gapped two-dimensional (2D) surface. Notice that, in this terminology, a first-order TI/TSC is just a conventional (strong) TI/TSC. Examples of higher-order TIs/TSCs discussed in recent works include three-dimensional (3D) insulators with hinge modes protected by rotation \cite{Fang17, Song17, Khalaf17}, mirror symmetry \cite{Langbehn17, Schindler17, Khalaf17, Benalcazar17}, or a combination of rotation and time-reversal \cite{Schindler17} as well as 2D insulators with corner modes protected by mirror symmetry \cite{Benalcazar16, Benalcazar17, Langbehn17}. Physical realizations of higher-order TIs have already been implemented in mechanical metamaterials \cite{Valerio18}, electronic circuits \cite{Imhof17}, and microwave resonators \cite{Peterson17} and proposed to exist in several materials \cite{Schindler17, Ezawa17, Ezawa18}. In addition, some of the systems considered in earlier works where a magnetic field is applied to a 3D TI \cite{Sitte12} or to ${}^3$He-B topological superfluid \cite{Volovik10} also fall in the category of higher-order TIs/TSCs.

In this work, we show that higher-order TIs/TSCs can be protected by inversion symmetry alone and provide a complete classification of inversion-protected $k$-th order TIs/TSCs in any spatial dimension. The crucial difference between inversion symmetry and the spatial symmetries considered in earlier works \cite{Benalcazar17, Schindler17, Langbehn17,Song17} lies in the fact that inversion does not leave any point on the surface invariant. This means that, in contrast to other symmetries such as rotation \cite{Fu11, Langbehn17} or mirror \cite{Fang17, Song17}, inversion-protected surface states cannot be observed by considering some symmetry-invariant plane on the surface. Instead, they can only be captured on particular sample geometries by considering the surface as a whole. Furthermore, inversion-protected surface states are not expected to be pinned to a particular edge or corner, but instead can be localized at any pair of corners or edges related to each other by inversion. This particular property allows for the possibility of controlling their position by applying inversion-preserving perturbations such as magnetic field, as we will show later. 

We will propose physical realizations for 3D second-order TIs, 3D second- and third-order TSCs and 2D second-order TSCs. The recipe for constructing these higher-order TIs/TSCs is to either apply a symmetry-breaking perturbation to a given strong (first-order) TI/TSC or to combine several strong TIs/TSCs such that the total strong index vanishes (while preserving the symmetry). The latter approach was used in constructing higher-order TIs in class AII in Refs.~\onlinecite{Fang17, Khalaf17} in which they were constructed by combining two time-reversal invariant 3D TIs in the presence of some spatial symmetries. The vanishing strong index implies that the surface can be gapped out by adding a mass term. The presence of spatial symmetries may, however, cast some global constraints on this mass term, forcing it to vanish on some subregion on the surface, e.g. a line or a set of points, on which the Hamiltonian remains gapless. The resulting system implements a higher-order TI/TSC. For example, an inversion-protected second-order 3D TI with or without TRS can be respectively constructed by combining two 3D strong TIs or by applying a magnetic field to a 3D strong TI. In both cases, the resulting system is trivial from the point of view of the 3D topology due to the vanishing strong index, but implements a second-order TI hosting propagating 1D hinge modes due to inversion.

It should be noted that the dimensionality of surface states, and consequently the ``order'' of a certain TCI, cannot be generally defined without specifying the geometry and boundary conditions of the sample. For example, a 3D TCI protected by mirror symmetry exhibits 2D gapless surface states on any surface plane that is left invariant under mirror symmetry \cite{Hsieh12, Tanaka12}. On the other hand, placing the same TCI on a sphere (with open boundary conditions in all directions) yields 1D hinge modes propagating along the circle where the mirror plane intersects the sphere \cite{Khalaf17}. This issue arises whenever the spatial symmetry protecting the phase leaves some subregion (point, line, etc.) on the surface invariant, as noted in Ref.~\onlinecite{Khalaf17} which circumvented the difficulty by not referring explicitly to order and studying all TCIs that host anomalous surface states on some surface (dubbed ``sTCI''). For inversion symmetry, however, this issue is not relevant since its action does not leave any point on the surface invariant (assuming the inversion center is in the bulk of the sample). Hence, in this work we will refer explicitly to the order $k$ of a higher-order TI/TSC to indicate the presence of $(d-k)$-dimensional surface states on {\it any} compact inversion-symmetric surface.

We would like to stress that the higher-order TIs/TSCs considered in this work are proper bulk topological phases that can only be trivialized by going through a phase transition closing the {\it bulk} gap. This means that the surface states considered here are anomalous, i.e. they cannot exist in a stand-alone lower-dimensional system. Our definition of the phase is consistent with the definition given in Refs.~\onlinecite{Schindler17, Khalaf17}, but differs from the definition used in Refs.~\onlinecite{Benalcazar17, Langbehn17}, where phases related by a {\it surface} phase transition were considered topologically distinct. 

We would also like to point out the relation between the phases considered here and the inversion-protected TCIs obtained using K-theory \cite{Lu14, Shiozaki14}. The K-theory approach classifies phases as equivalence classes of bulk Hamiltonians that can be (stably) deformed into each other without closing the bulk gap or breaking the protecting symmetries. This means that the inversion-protected higher-order TIs/TSCs considered here are stable K-theory phases. On the other hand, it is easy to show that some of the K-theory phases, e.g. those corresponding to distinct atomic limits, do not possess any types of surface states and can thus be considered trivial from the perspective of surface states. Our classification scheme considers two phases the same if they differ by the addition of a phase with gapped surface even if they cannot be adiabatically connected. Thus, the set of phases we obtain here is equivalent to the K-theory phases\cite{Lu14, Shiozaki14} modulo those without surface states.

This paper is structured as follows. We start by summarizing the main argument used for understanding and classifying inversion-protected higher-order TIs/TSCs in Sec.~\ref{Summary}. We show that higher-order TIs/TSCs can generally be understood as ``globally irremovable surface topological defects'', which is then used to derive a necessary condition for the existence of inversion-protected $k$-th order TI/TSC in any given dimension. In Sec.~\ref{Realizations}, we propose physical realizations for 3D second-order TIs, 3D second- and third-order TSCs and 2D second-order TSCs. We discuss how these phases can be constructed by combining conventional TIs/TSCs or applying symmetry-breaking perturbations (e.g. magnetic field) to them and provide the pattern of surface states expected in each case. In Sec.~\ref{Layer}, we make use of the layer construction introduced in Refs.~\onlinecite{Hermele17, Huang17} to provide a full classification of inversion-protected $k$-th order TI/TSC in any dimension leading to Table \ref{Classification}. Afterwards, we provide a minimal Dirac model for the inversion-symmetric higher-order TIs/TSCs in any dimension in Sec.~\ref{Dirac}. We close by making several concluding remarks regarding the stability of the phase against symmetry-breaking perturbations and the generalization to other spatial symmetries in Sec.~\ref{Discussion}.

\section{Summary of the argument}
\label{Summary}
We begin by summarizing our main argument for the construction and complete classification of inversion-protected higher-order TIs/TSCs. Following Ref.~\onlinecite{Khalaf17}, we think of the surface states of higher-order TIs/TSCs as globally irremovable topological defects. Recall that a topological defect corresponds to a region in space, e.g. a domain wall, where some parameter in the Hamiltonian is changed such that it hosts zero energy states. Being topological means that the zero energy states are robust against any symmetry-preserving perturbations, but they can generally be moved freely. The simplest example of a topological defect is the surface of a strong TI/TSC which can be thought of as a domain wall between the bulk topological Hamiltonian and the trivial vacuum outside. More interesting examples include vortices in 2D $p_x + i p_y$ superconductors, which host Majorana zero modes \cite{Read00, Ivanov01}, or dislocations in layered topological insulators \cite{Teo10}. 

The local stability of a topological defect depends only on the dimension of the defect (which is the difference between the spatial dimension and the co-dimension of the defect) and the presence or absence of the local symmetries $\T$, $\P$, and $\S$ \cite{Teo10}. Using this knowledge, one can immediately deduce the classification of stable topological defects of a certain dimension from the classification of strong TIs/TSCs in one dimension higher. The resulting classification table in any dimension for the ten AZ classes was given in Ref.~\onlinecite{Teo10} and it is identical to the classification table of TIs/TSCs \cite{Kitaev09, Schnyder09, Ryu10} with the dimension shifted by 1. Similar to the classification of TIs/TSCs, topological defects of any given dimension are stable in five symmetry classes, three of which can host an arbitrary number of gapless states in the defect ($\Z$ defects), while two can host at most one gapless state ($\Z_2$ defects). 

In general, a topological defect on an otherwise gapped compact surface can always be removed \footnote{Whenever we refer to a compact surface in this work, we implicitly assume it has genus 0, e.g. a sphere.}. For instance, on the surface of a sphere, a line defect can always be deformed to a point. Similarly, point defects have to occur in pairs which can be annihilated by bringing them together. Thus, despite their local stability, topological defects whose dimension is lower than that of the surface are globally unstable. Spatial symmetries, however, can make it impossible to remove these defects without breaking the symmetry. In the case of inversion, this follows from the fact that no point on the surface is left invariant by inversion, thus an inversion-symmetric topological defect can never be deformed to a point. For example, inversion forces a pair of point defects to be located at two inversion-related (antipodal points) which can never be brought together without breaking inversion. Likewise, line defects will be confined to inversion-symmetric curves which cannot be deformed to a point.

\begin{figure}
\center
\includegraphics[width=0.95\columnwidth]{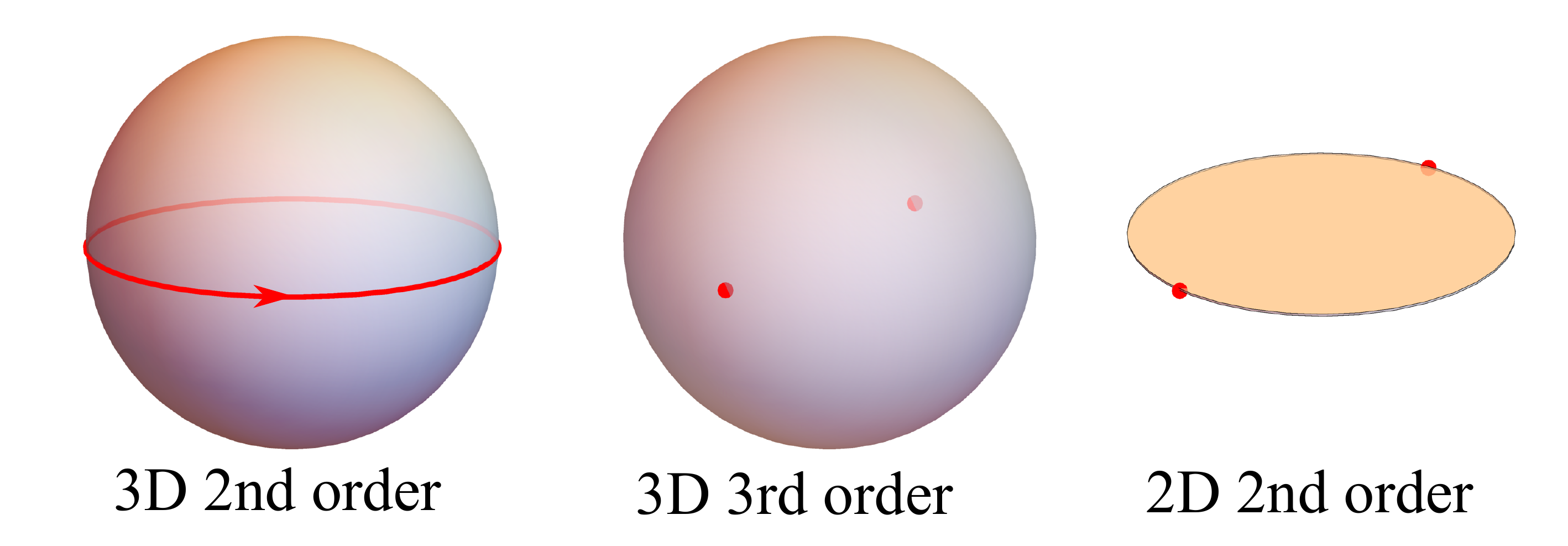}
\caption{Surface states of inversion-protected higher-order TI/TSC in two and three dimensions.}
\label{HOTI}
\end{figure}

The correspondence between inversion-protected surface states and topological defects can be established by noting that inversion acts non-locally on the surface by mapping different points to each other, thus the only symmetries which can stabilize the gapless states locally are $\T$, $\P$, and $\S$. As a result, the surface states of an inversion-protected $k$-th order topological insulator in a given symmetry class in $d$ dimensions are locally identical to $(d-k)$-dimensional topological defects in the same class and are thus only stable if these defects are stable \footnote{A similar argument can be made for point group symmetries other than inversion. In that case, however, we need to be careful when we consider points/lines which are left invariant by the symmetry, where the spatial symmetry acts as an onsite unitary symmetry, possibly leading to a different effective symmetry class at these invariant regions.}. Using the notation of Ref.~\onlinecite{Teo10}, which labels the complex AZ classes by a $\Z_2$ variable $s_c = 0,1$ for classes A and AIII, respectively, and the real AZ classes by a $\Z_8$ variable $s_r = 0,\dots,7$ for classes AI, BDI, $\dots$, CI, respectively (cf.~Table \ref{Classification}), this conditions implies that an inversion-protected $k$-th order TI/TSC is possible in $d$ dimensions when $s_c - d + k - 1= 0 \mod 2$ for complex classes or when $s_r - d + k - 1= 0,1,2,4 \mod 8$ for real classes.

\begin{figure}[t]
\center
\includegraphics[width=0.95\columnwidth]{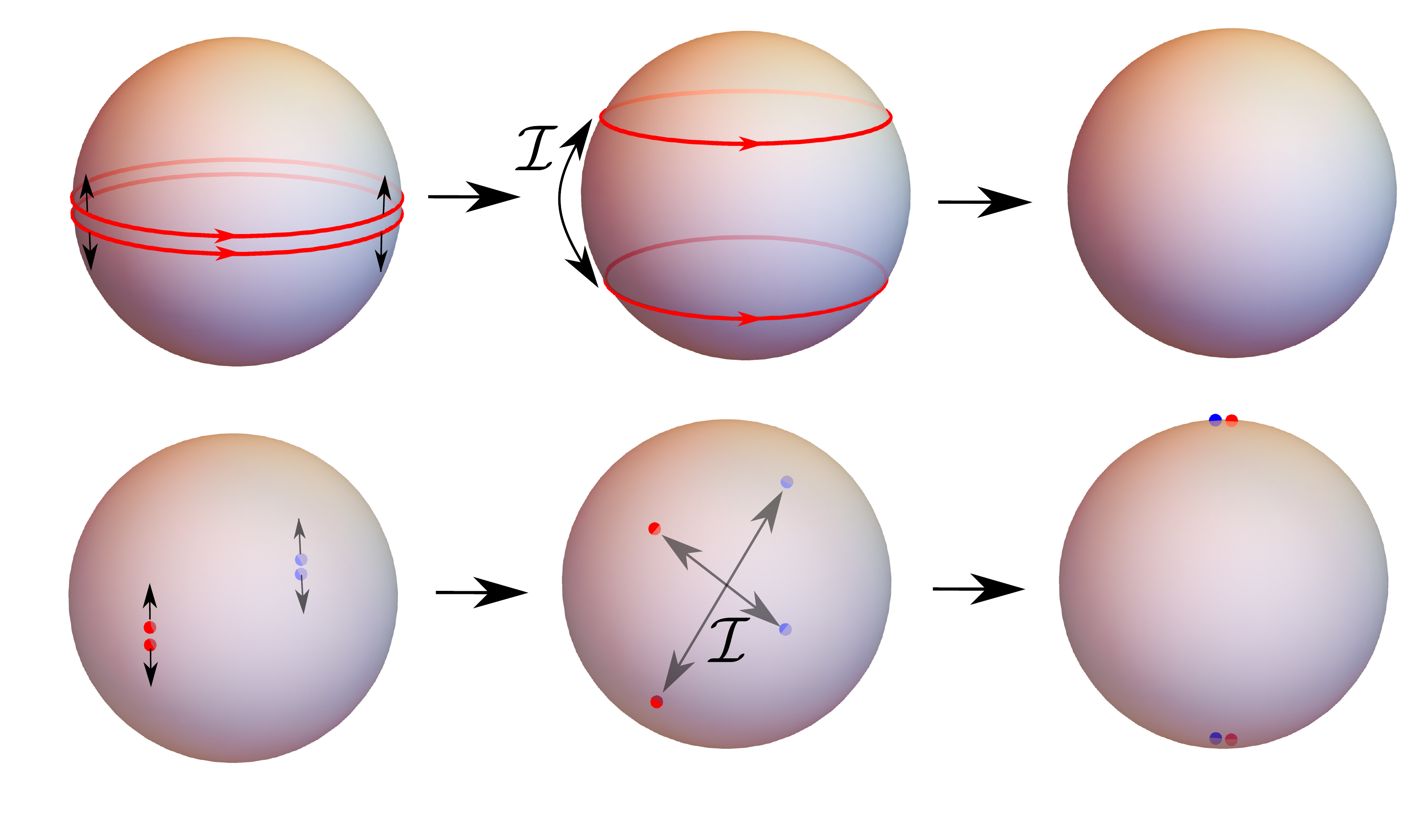}
\caption{The upper panel illustrates how two chiral modes can be removed without breaking inversion symmetry. The lower panel illustrates how a pair of 0D topological defects (each defects consists of two zero modes at two inversion related points with opposite chirality) can be removed by moving states of opposite chirality towards each other and annihilating them. As a result, the classification of inversion-protected higher-order TIs/TSCs is always $\Z_2$ regardless of the classification of the defect.}
\label{Z2}
\end{figure}

One specific feature of inversion-protected higher-order TIs/TSCs is that they always have a $\Z_2$ classification regardless of the underlying classification of the defects (This means that the stability of $(d-k)$-dimensional topological defects is a necessary but not sufficient condition for the existence of an inversion-protected $k$-th order TI/TSC in $d$ dimensions). The reason for this is that a pair of inversion-symmetric topological defects (notice that an inversion-symmetric 0D defect consists of two points) can each be deformed to a point without breaking inversion. This is achieved by ensuring they are related to each other by inversion at every stage of the process. This process is visually illustrated in Fig.~\ref{Z2} for zero- and one-dimensional defects on the surface of a sphere. The upper panel illustrates how two chiral modes living on a 1D curve on the surface, for example in class A, can be removed while preserving inversion. This process is equivalent to adding a 2D quantum Hall layer on the upper hemisphere and its inverted copy on the lower one, which should have no influence on the bulk phase (a similar argument was used in Ref.~\onlinecite{Schindler17}). The lower panel illustrates how this can be done for $\Z$-type 0D defects, for example in class BDI. In this case, the zero modes can be assigned a definite chirality such that the total chirality over the whole surface vanishes (in a 1D BDI wire, chirality distinguishes the zero mode at the two endpoints such that the two modes at the same endpoint do not hybridize but two modes belonging to different endpoints can gap out by hybridizing). Defects of opposite chirality can be brought together and annihilated as shown in the figure. This is equivalent to adding a BDI Majorana chain and its inverted copy to the surface.

The complete classification of inversion-protected $k$-th order TI/TSC in a given symmetry class in a given dimension $d$ is given in Table \ref{Classification}. It should be noted that only inversion-protected phases (phases which are trivialized upon breaking inversion) are considered. This automatically excludes standard (first-order) TIs/TSCs. As anticipated, the classification depends only on the dimensionality of the surface states $d-k$ and is always either 0 or $\Z_2$. Recall that the action of $\T$, $\P$, $\S$ (when present) and $\I$ on the Hamiltonian is given by
\begin{gather}
\T^{-1} \H_\bk^* \T = \H_{-\bk}, \quad \P^{-1} \H_\bk^* \P = -\H_{-\bk} \nonumber \\
\S^{-1} \H_\bk \S = - \H_\bk, \quad \I^{-1} \H_\bk \I = \H_{-\bk}.
\label{Sym}
\end{gather}
The last piece of information needed to determine whether a higher-order TIs/TSCs is possible in a given dimension and symmetry class is the commutation/anticommutation relations between inversion and the local symmetries $\T$, $\P$, and $\S$. This information is provided in Table \ref{Classification} via the superscripts for $\Z_2$ , which indicate commutation ($+$) or anticommutation ($-$) between inversion symmetry and $\T$, $\P$, or $\S$ (when present), such that a higher-order TI/TSC is possible. When the three symmetries $\T$, $\P$, or $\S$ are present, we specify the commutation/anticommutation sign for time-reversal followed by that for particle-hole (we leave out the sign for chiral symmetry which is fixed by these two). The signs placed between brackets indicate that for these commutation relations, a $k$-th order TI/TSC is only possible up to a finite order $k_m=2$ or $3$, which is indicated by the corresponding subscript. For example, the entry $\Z_2^{+-,(--)_2}$ for class DIII when $d-k+1=2 \mod 8$ means that it is possible to have a $k$-th order phase in $d$ dimensions for any $k$ when $\I$ commutes with $\T$ and anticommutes $\P$, but it is only possible for orders $k \leq 2$ when $\I$ anticommutes with both $\T$ and $\P$. For the classification in Table \ref{Classification}, inversion is assumed to square to $+1$. The case of $\I^2 = -1$, which is relevant for example for  inversion symmetry for spinful electrons in strictly 2D systems, is obtained by replacing $\I \rightarrow i \I$. This replacement changes the commutation/anticommutation sign for the anti-unitary symmetries $\T$ and $\P$ but leaves the sign for $\S$ unchanged. The derivation of the commutation/anticommutation constraints given in the Table will be provided in Sec.~\ref{Layer} by means of a layer construction.

\begin{center}
\begin{table*}[t]
\caption{{\bf Classification of higher-order TIs/TSCs protected by inversion symmetry}. The first column $s$ indicates the Bott label for the complex and real classes, second column the AZ (or Cartan) label, and the next three columns indicate the presence ($\pm 1$) or absence (0) of time-reversal and particle-hole symmetries distinguishing the cases of $\T^2, \P^2 = \pm 1$ as well as the presence (1) or absence (0) of chiral symmetry $\S = \T \P$. The next 8 columns give the classification of inversion-protected $k$-th order TI/TSC in $d$ dimensions, which depends only on $\delta = d - k + 1$ ($k=1$ corresponds to standard TIs/TSCs). The superscripts to the $\Z_2$ entries indicate the commutation properties of the inversion symmetry with the local symmetries $\T$, $\C$ and $\S$ (when present) for which the non-trivial phase is possible. For class AIII, the superscript $\sigma_S = \pm$ is defined as $\S^{-1} \I \S = \sigma_\S \I$. For real classes with a single antiunitary symmetry $\A = \T, \P$ (AI, D, AII, C), the superscript $\sigma_\A = \pm$ is similarly defined as $\A^{-1} \I \A = \sigma_\A \I$. For real classes with both $\T$ and $\P$ symmetries (BDI, DIII, CII, CI), the two superscripts  $\sigma_{\T,\P} = \pm$ are defined as $\T^{-1} \I \T = \sigma_\T \I$ and $\P^{-1} \I \P = \sigma_\P \I$, respectively ($\sigma_\T$ always occurs first). The superscripts between brackets indicate the commutation/anti-commutation signatures for which a higher-order TI/TSC is only possible up to a certain value of the order $k = k_m$, with $k_m = 2,3$ indicated as a subscript for the bracket.}
\bgroup
\setlength{\tabcolsep}{0 em}
\setlength\extrarowheight{0.25em}
\begin{tabular}{c|c|ccc|cccccccc}
\hline \hline
\multicolumn{5}{c|}{Symmetry} & \multicolumn{8}{c}{$\delta = d- k + 1$} \\
\hline
\rule{0pt}{2.5ex} 
$\,s\, $ & $\,$ AZ $\,$ & $\, \T^2\,$ & $\,\P^2\,$ & $\,\S^2\,$ & 0 & 1 & 2 & 3 & 4 & 5 & 6 & 7\\
\hline
\rule{0pt}{2ex} 
0 & A & 0 & 0 & 0 & $\Z_2$ & 0 & $\Z_2$ & 0 & $\Z_2$ & 0 & $\Z_2$ & 0\\
1 & AIII & 0 & 0 & 1 & 0 & $\Z_2^{-}$ & 0 & $\Z_2^{-}$ & 0 & $\Z_2^{-}$ & 0 & $\Z_2^{-}$\\
\hline
\rule{0pt}{3ex} 
0 & AI & 1 & 0 & 0  & $\Z_2^{+}$ & 0 & 0 & 0 & $\Z_2^{+,(-)_3}$ & 0 & $\Z_2^{+,(-)_2}$ & $\Z_2^{+}$ \\
1 & BDI & 1 & 1 & 1 & $\,\Z_2^{+-,(++)_2}$ & $\Z_2^{+-}$ & 0 & 0 & 0 & $\Z_2^{+-,(-+)_3}$ & 0 & $\Z_2^{+-,(++,-+)_2}$ \\ 
2 & D & 0 & 1 & 0 & $\Z_2^{-,(+)_2}$ & $\Z_2^{-}$ & $\Z_2^{-}$ & 0 & 0 & 0 & $\Z_2^{-,(+)_3}$ & 0  \\
3 & DIII & $-1$ & 1 & 1 & 0& $\Z_2^{+-,(--,-+)_2}\,\,$ & $\Z_2^{+-,(--)_2}$ & $\Z_2^{+-}$ & 0 & 0 & 0 & $\Z_2^{+-,(-+)_3}$  \\ 
4 & AII &$-1$ & 0 & 0 & $\Z_2^{+,(-)_3}$ & 0 & $\Z_2^{+,(-)_2}$ & $\Z_2^{+}$ & $\Z_2^{+}$ & 0 & 0 & 0  \\
5 & CII & $-1$ & $-1$ & 1 & 0 & $\Z_2^{+-,(-+)_3}$ & 0 & $\Z_2^{+-,(++,-+)_2}\,\,$ & $\Z_2^{+-,(++)_2}$ & $\Z_2^{+-}$ & 0 & 0  \\
6 & C & 0 & $-1$ & 0 & 0 & 0 & $\Z_2^{-,(+)_3}$ & 0 & $\Z_2^{-,(+)_2}$ & $\Z_2^{-}$ & $\Z_2^{-}$ & 0    \\
7 & CI & 1 & $-1$ & 1  & 0 & 0 & 0 & $\Z_2^{+-,(-+)_3}$ & 0 & $\Z_2^{+-,(--,-+)_2}\,\,$ & $\Z_2^{+-,(--)_2}$ & $\Z_2^{+-}$    \\
\hline \hline
\end{tabular}
\egroup
\label{Classification}
\end{table*}
\end{center}

\section{Physical realizations}
\label{Realizations}


\subsection{Inversion symmetry in Bogoliubov-de Gennes Hamiltonians}
\label{BdG}
Before discussing concrete physical realization for higher-order TIs/TSCs protected by inversion, it is instructive to review how inversion symmetry is implemented in Bogoliubov-de Gennes (BdG) Hamiltonians, which describe superconductors. A BdG Hamiltonian is given by
\beq
H = \frac{1}{2} \sum_{\bk} (\psi^{\dagger}_{\bk}, \psi_{-\bk}) \H_\bk \left(\begin{array}{c} \psi_{\bk} \\ \psi^{\dagger}_{-\bk}\end{array}\right), \,\, \H_\bk =  \left(\begin{array}{cc} \Xi_\bk & \Delta_\bk\\\Delta_\bk^{\dagger} & -\Xi_{-\bk}^T\end{array}\right),
\eeq
with $\psi_{\bk} = (c_{\bk,\uparrow},c_{\bk,\downarrow})$, $\Xi_\bk$ representing the single particle part and $\Delta_\bk$ the pairing or gap function. The Hamiltonian $\H_\bk$ satisfies the PHS $\tau_x \H^*_{\bk} \tau_x = -\H_{-\bk}$ with $\boldsymbol \tau$ representing the Pauli matrices in the particle-hole (Nambu) space. For a centrosymmetric superconductor, the single particle part $\Xi_\bk$ is invariant under inversion $\I^{-1} \Xi_\bk \I = \Xi_{-\bk}$. For the gap function $\Delta_\bk$, the two possibilities $\Delta_{-\bk} = \pm \Delta_\bk$ are consistent with inversion symmetry and they correspond to even ($+$) or odd ($-$) parity superconductors. It should be noted that in a simple two band model without spin-orbit coupling, even (odd) parity corresponds to a singlet (triplet) superconductor, but this relation does not hold in more complicated models \cite{Fu10, Sato10, Ando15, Sato17}. Inversion symmetry $\I$ can be promoted to a symmetry $\I^\tau$ of the full BdG Hamiltonian by extending it to the Nambu space as
\beq
\label{IParity}
\I^\tau = 
\begin{cases}
\I \otimes \tau_0, &: \Delta_{-\bk} = \Delta_\bk, \\
\I \otimes \tau_z, &: \Delta_{-\bk} = -\Delta_\bk, 
\end{cases}
\eeq
which commutes (anticommutes) with PHS for even (odd) parity when $\I^2 = 1$. We note that the form of inversion symmetry (\ref{IParity}) in odd parity superconductors was previously derived in Refs.~\onlinecite{Fu10, Sato10}, which concluded that odd parity is required to obtain a time-reversal symmetric TSC in 3D. According to Table \ref{Classification}, most of the physically relevant cases for higher-order TIs/TSCs, e.g. second- or third-order class D in two or three dimensions, can only be achieved when inversion anti-commutes with PHS, which, as we have just seen, corresponds to odd-parity superconductors. 

\subsection{Higher-order three-dimensional topological insulators}
\label{3DTI}
In this section, we consider examples of higher-order TIs in 3D. According to Table \ref{Classification}, third-order TIs (classes A, AI or AII) are impossible in 3D, whereas second-order ones are possible in 3D in the presence or absence of spinful TRS $\T^2=-1$ corresponding to classes AII or A, respectively. In class A, this can be implemented by applying magnetic field (which breaks time-reversal but preserves inversion) to a time-reversal invariant 3D TI. To see this, let us consider the Hamiltonian for a 3D TI
\begin{multline}
\label{HTI}
\H = (\sin k_x \sigma_x + \sin k_y \sigma_y + \sin k_z \sigma_z) \tau_x \\ - (3-\lambda-\cos k_x - \cos k_y - \cos k_z) \sigma_0 \tau_z, \quad \lambda=1.
\end{multline}
Here, $\boldsymbol \tau$ and $\boldsymbol \sigma$ denote the Pauli matrices in the orbital and spin spaces, respectively (notice that we use $\boldsymbol \tau$ here to indicate orbital space unlike Secs.~\ref{BdG}, \ref{3DTSC}, and \ref{2DTSC}, where it indicates Nambu space).
The Hamiltonian (\ref{HTI}) is invariant under inversion and time-reversal symmetries given respectively by $\I = \sigma_0 \tau_z$ and $\T = i \sigma_y \tau_0 \K$ ($\K$ here indicates complex conjugation as usual).

Next, we place the system on some compact inversion-symmetric manifold in 3D (open boundary conditions in all directions). The derivation of the surface theory follows the standard procedure \cite{Jackiw76} by introducing a spatial dependence in the mass parameter $\lambda \rightarrow \lambda_t$, where $t$ denotes the distance from the surface along the perpendicular direction. The surface is defined such that $\lambda_0 = 0$ and $\lambda_t \rightarrow \pm 1$ quickly away from the surface with the interior (exterior) of the sample corresponding to $+1$ ($-1$). We next linearize the Hamiltonian by expanding in small momenta around the $\Gamma$ point and consider it close to the surface
\beq
\label{HTIcont}
\H =  \bsigma \cdot \bk_S \tau_x + [-i \bsigma \cdot \bn_\br \tau_x \partial_t + \lambda_t \sigma_0 \tau_z].
\eeq
Here, $\bn_\br$ indicates the normal to the surface at point $\br$, satisfying $\bn_{-\br} = -\bn_\br$ and $\bk_S$ is the momentum parallel to the surface (we assume the surface is a smooth manifold so that the surface momentum at a given by point can be defined as a vector in the tangent space at this point). We now look for the eigenstates $\psi_{t,\bk_S}$ of (\ref{HTIcont}) that are localized close to the surface. These are fixed by the requirements that they decay fast enough away from the surface and are annihilated by the term between square brackets in (\ref{HTIcont}) to be
\beq
\psi_{t,\bk_S} = e^{\int_0^t dt' \lambda_t} P \psi_{\bk_S}, \quad P = \frac{1}{2}(1 - \bn_\br \cdot \bsigma \, \tau_y).
\label{Pr}
\eeq
Here, $P$ denotes the surface projection operator \cite{Khalaf17}, which can be simplified by rotating to a basis where it is diagonal using 
\beq
\label{Ur}
U_\br = \exp\left(i \frac{\pi}{4} \tau_x \, \bn_\br \cdot \bsigma \right) \quad \Rightarrow U_\br^{\dagger} P U_\br = \frac{1}{2}(1 - \tau_z).
\eeq
We can now introduce the 4$\times$2 matrix $p = (0, \sigma_0)^T$ which acts on a 4$\times$4 matrix in the $\tau$ and $\sigma$ space to pick up the 2$\times$2 block corresponding to the non-zero eigenvalue of the rotated projector. The surface Hamiltonian can then be explicitly obtained as
\begin{align}
\label{HSAII}
h &= p^T U_\br^{\dagger} \tau_x \bsigma \cdot \bk U_\br p =
p^T i \tau_0 (\bsigma \cdot \bk_S)(\bn \cdot \bsigma)] p \nonumber\\
&= -(\bk_S \times \bn) \cdot \bsigma = -(\bk \times \bn) \cdot \bsigma,
\end{align}
where we used the fact that $\bn \times \bn = 0$ in the last line to replace the surface momentum $\bk_S$ by the total momentum $\bk = -i \bnabla$. The symmetry action on the surface can be likewise deduced from its bulk action by the use of the projector (\ref{Pr}) and basis rotation (\ref{Ur}). Time-reversal symmetry acts on the surface degrees of freedom as
\beq
\T_S = p^T U_\br^{\dagger} i\sigma_y \tau_0 \K U_\br p = i \sigma_y \K,
\eeq
which is the same as its action in the bulk. Inversion symmetry acts  by mapping $\br$ to $-\br$ and flipping the momentum $\bk$ and is represented on the surface by 
\beq
\I_S = p^T U_\br^{\dagger} \sigma_0 \tau_z U_{-\br} p = -\sigma_0.
\eeq
We notice that, in principle, the symmetry action on the surface could depend on the point $\br$ although this is not the case here.

We now allow for TRS-breaking mass terms that preserve inversion, which can be achieved by applying a magnetic field. The only possible such term has the form $m_\br \bn \cdot \bsigma$ (this can be obtained as the surface projection of the bulk mass term $m_\br \tau_y$). Inversion symmetry requires $m_\br$ to satisfy $m_{-\br} = -m_\br$. As a result, this mass term has to vanish as we go between any point and its image under inversion along any curve, which in turn implies the existence of a 1D inversion-symmetric curve on which the mass term vanishes. Such 1D curve will host a chiral gapless mode analogous to the edge modes in a quantum Hall sample. The TRS-breaking mass term $m_\br \bn \cdot \bsigma$ can be identified as a Zeeman term resulting from applying a uniform magnetic field, with $m_\br$ being proportional to the field component normal to the surface. The curve along which the mass term vanishes corresponds to the region where the magnetic field is tangent to the surface and thus can be gauged away. Such a phase was studied before in Ref.~\onlinecite{Sitte12}.

\begin{figure}
\center
\includegraphics[width=0.75\columnwidth]{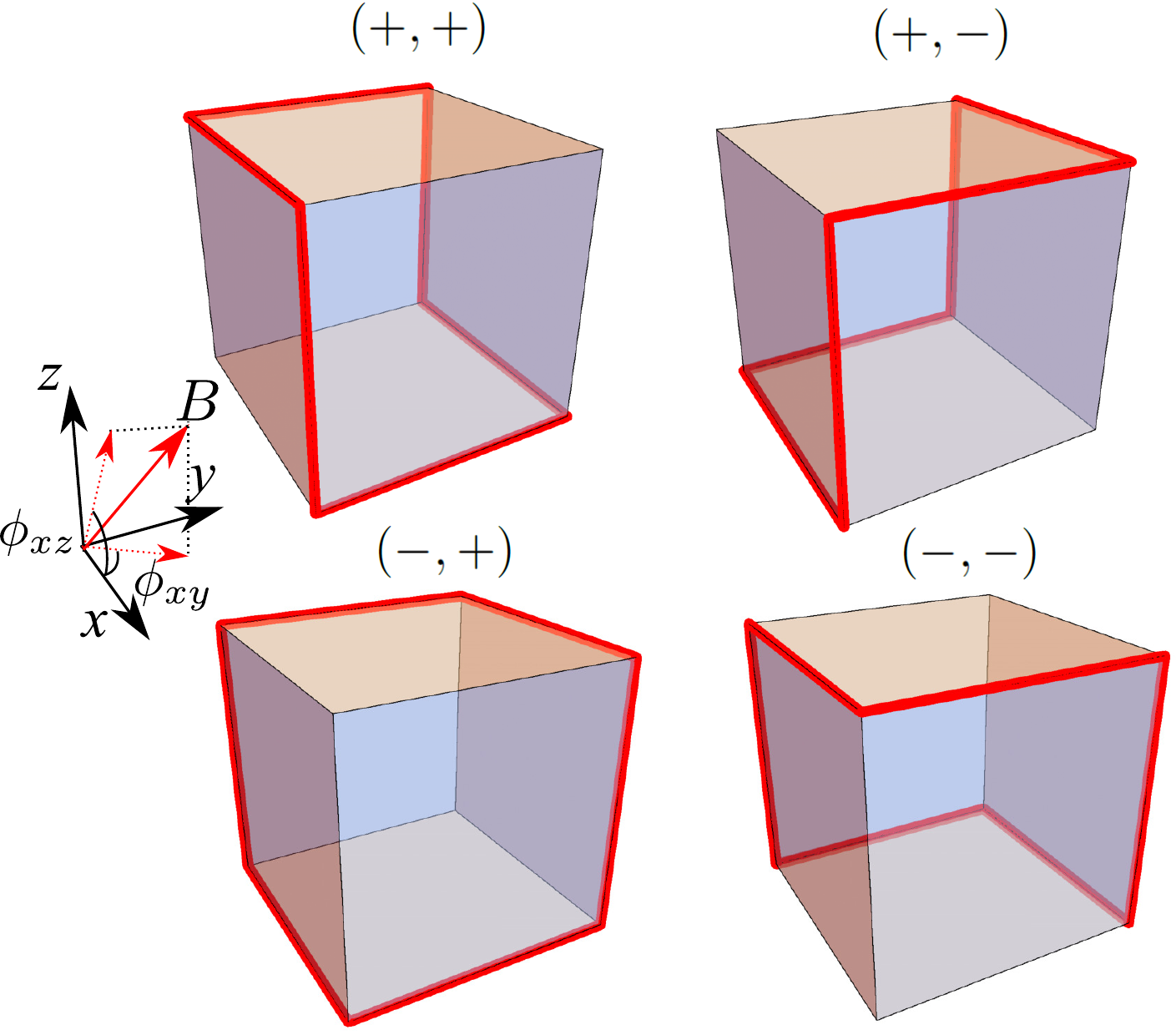}
\caption{Illustration of the pattern of surface states obtained by applying a magnetic field to a cubic sample of a time-reversal symmetric 3D TI. The four possible patterns correspond to the four possible values $(\sgn \tan \phi_{xy}, \sgn \tan \phi_{xz}) = (\pm, \pm)$, where $\phi_{xy}$ and $\phi_{xz}$ correspond to the angle between the $x$-axis and the projection of the field on the $x$-$y$ plane and the $x$-$z$ plane, respectively.}
\label{Cube}
\end{figure}

In a cubic sample, the mass term is not expected to change along any face of the cube, thus we expect the chiral 1D mode to propagate along the edges or ``hinges'' of the sample. When the applied field is not parallel to any face, there are four possible configurations of inversion-symmetric edges where the modes can propagate, shown in Fig.~\ref{Cube} (for each configuration, the chiral mode can propagate in either direction), depending on the direction of the applied field. The Hamiltonian is expected to be gapless at edges where the field is tangent to the surface so that it can be gauged away. This leads to the dependence of the edge modes on the applied field schematically illustrated in Fig.~\ref{Cube}, where the four possible configurations correspond to the values $(\sgn \tan \phi_{xy}, \sgn \tan \phi_{xz}) = (\pm, \pm)$. Here, $\phi_{xy}$ and $\phi_{xz}$ correspond to the angle between the $x$-axis and the projection of the field on the $x$-$y$ plane and the $x$-$z$ plane, respectively.

A time-reversal symmetric variant of this phase can be achieved by ``stacking'' two 3D TIs together as was shown in Refs.~\onlinecite{Po17, Khalaf17}. This can be seen by observing that a time-reversal-symmetric mass term can be added in a system consisting of two copies of the surface Hamiltonian (\ref{HSAII}), which is consistent with the fact that a 3D TI have a $\Z_2$ classification. Such mass term has the form $m_\br \bn \cdot \bsigma \mu_y$ ($\boldsymbol \mu$ are the Pauli matrices in the space of the two copies) and, in the presence of inversion, it will necessarily satisfy $m_{-\br} = -m_\br$. As a result, it will vanish along an inversion-symmetric 1D curve on the surface, which will host a propagating helical gapless mode similar to the edge of a quantum spin-Hall sample. This state can be thought of as a sum of the broken TRS case considered earlier and its time-reversed copy \cite{Fang17}.

\subsection{Higher-order three-dimensional topological superconductors}
\label{3DTSC}
We now consider examples of higher-order TSCs in 3D. According to Table \ref{Classification}, it is possible to have higher order TSCs with 1D (second-order) or 0D (third-order) surface states in triplet superconductors ($\P^2 = +1$) with (class DIII) or without (class D) TRS. Our approach to constructing such phases follows the previous section by either combining two 3D TSCs such that their strong 3D topological invariant vanishes or applying a TRS-breaking perturbation to a time-reversal-symmetric TSC. In both cases, the surface can be gapped out by writing a mass term which will be forced by inversion to vanish on a line or a pair of points on the surface. Our starting point is a topological superconductor of class DIII (spinful TRS), which can be thought of as the superconducting analog of a 3D strong TI but which differs in the fact that is has a $\Z$, rather than $\Z_2$, classification. 

A topologically non-trivial state $\nu = \pm 1$ in class DIII describes the $B$-phase in liquid ${}^3 {\rm He}$ \cite{Volovik03, Schnyder09} with $\Delta_\bk = \bk \cdot \bsigma (i \sigma_y)$. The low energy effective Hamiltonian can be obtained from the BdG  Hamiltonian by substituting $\Delta_\bk = \bk \cdot \bsigma (i \sigma_y)$ and $\Xi_\bk = \lambda$ leading to
\beq
\label{HDIII3}
\H_\xi = \xi(-k_x \sigma_z \tau_x - k_y \tau_y + k_z \sigma_x \tau_x) + \lambda \sigma_0 \tau_z, \qquad \xi=\pm 1.
\eeq
Here, $\boldsymbol \tau$ indicates the Pauli matrices in the Nambu space as in Sec.~\ref{BdG} and $\boldsymbol \sigma$ indicates, as usual, the spin Pauli matrices. The Hamiltonian (\ref{HDIII3}) is invariant under TRS $\T = \sigma_y \tau_0 \K$, PHS $\P = \sigma_0 \tau_x \K$, and consequently the chiral symmetry $\S = \sigma_y \tau_x$. It is also invariant under inversion symmetry implemented as $\I = \sigma_0 \tau_z$, which commutes with $\T$, but anti-commues with $\P$. This is consistent with Table \ref{Classification} and also with the discussion of Sec.~\ref{BdG} about inversion symmetry in odd-parity superconductors. Here, $\xi = \pm 1$ corresponds to phases with different topological indices $\nu =\pm 1$ and $\lambda$ gives the negative of the chemical potential at $\bk = 0$ (we assume that the single-particle dispersion depends weakly on $\bk$ close to $\bk = 0$).

We will find it more convenient to perform the basis rotation 
\beq
\H_\xi \rightarrow V^{\dagger} \H_\xi V, \qquad V = e^{i \frac{\pi}{4} \sigma_y (\tau_z - \tau_0) }.
\eeq
In the new basis, the Hamiltonian (\ref{HDIII3}) becomes
\beq
\label{Hxi}
\H_\xi = \xi \bk \cdot \bsigma \tau_x + \lambda \sigma_0 \tau_z,
\eeq
which has the same form as (\ref{HTIcont}). Inversion and TRS are unaffected by the change of basis, while PHS and chiral symmetry are given in the new basis by $\P = \sigma_y \tau_y \K$ and $\S = \sigma_0 \tau_y$, respectively.

The surface theory is derived as in Sec.~\ref{3DTI} using the rotation $U_\br$ defined in (\ref{Ur}) followed by the projection 
\beq
p_\xi = \begin{cases} (0, \sigma_0)^T &: \xi = +, \\
(\sigma_0,0)^T &: \xi = -, \end{cases}
\eeq
leading to the surface Hamiltonian
\beq
\label{hxi}
h_{\xi} = -\xi (\bk \times \bn_\br) \cdot \bsigma.
\eeq
One major subtlety in this case is the following: whereas the surface implementation of TRS and inversion, given respectively by $\T_S = \sigma_y \K$ and $\I_S = -\xi \sigma_0$, is independent on $\br$, the implementation of PHS (and consequently chiral symmetry) does depend on the point $\br$ on the surface. It is easier to see this from the surface form of $\S$ which is
\beq
\S_S = p^T_\xi U^\dagger_\br  \S U_\br p_\xi = p^T_\xi \S U_\br^2 p_\xi = p^T_\xi \bn \cdot \bsigma \tau_z p_\xi = -\xi \bn \cdot \bsigma
\eeq
It follows that $\P_S = -\xi \bn \cdot \bsigma \sigma_y \K$. The requirement of anticommutation with $\S_S$ forbids any possible mass term even when we consider several copies of (\ref{hxi}) with the same $\xi$, reflecting the $\Z$ classification. 



Let us now combine two copies of the Hamiltonian (\ref{Hxi}) with opposite topological index $\H_+ \oplus \H_-$ and consider the surface theory, which we can write more explicitly by introducing the Pauli matrices $\boldsymbol \gamma$ in the space of the two copies as
\beq
\label{hSDIII}
h = -(\bk \times \bn_\br) \cdot \bsigma \gamma_z.
\eeq
Here, TRS, PHS, and inversion are implemented as $\T_S = \sigma_y \K$, $\P_S = -\gamma_z \bn \cdot \bsigma \sigma_y \K$, and $\I = -\gamma_z$, respectively. Due to the vanishing strong index, we can write a mass term $m_\br \gamma_x$ which is invariant under TRS and PHS. Inversion implies $m_{-\br} = -m_\br$ leading to a single helical 1D Majorana surface mode propagating along the inversion-symmetric 1D curve where the mass vanishes. This phase is the superconducting analog of the doubled strong TI discussed in the previous section (also in Refs.~\onlinecite{Khalaf17, Po17, Fang17}) and it implements a second-order TSC in class DIII. Similarly, we can implement a second-order TSC in class D by applying a magnetic field to the surface Hamiltonian of a single copy of the DIII Hamiltonian (\ref{hxi}). In this case, the only mass term consistent with preserved PHS and broken TRS is $m_\br \bn \cdot \bsigma$. This mass term is again required by inversion symmetry to vanish along a 1D inversion-symmetric curve which will host a propagating chiral Majorana mode. This phase can be realized by applying a magnetic field to topological superfluid ${}^3$He-B and it was previously studied in Ref.~\onlinecite{Volovik10}.

\begin{figure}
\center
\includegraphics[width=\columnwidth]{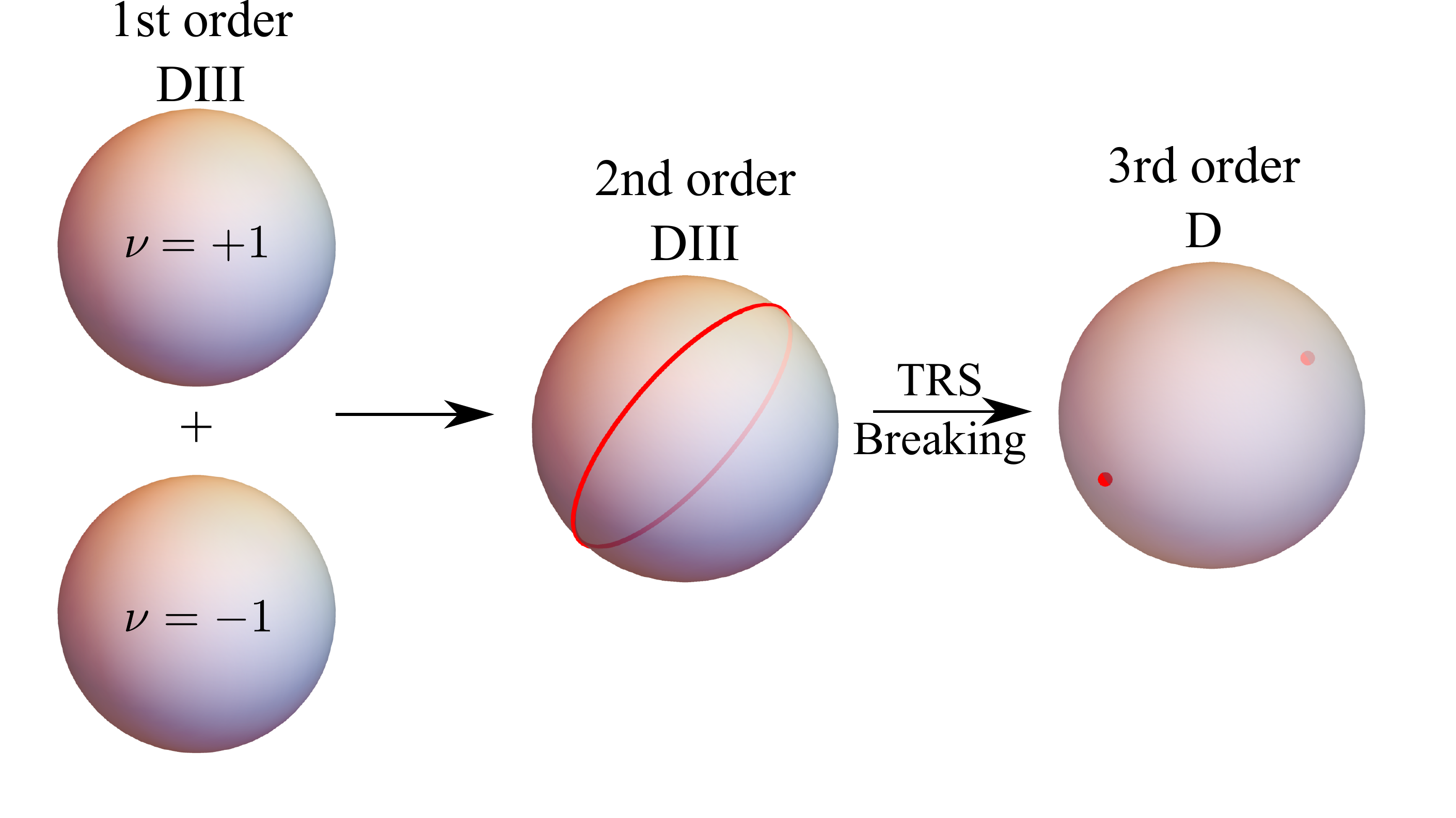}
\caption{Schematic illustration of the construction of a third-order TSC with two corner states in 3D. The first stage involves combining two (first-order) 3D time-reversal symmetric TSCs (class DIII) with opposite topological index $\nu = \pm 1$ leading to a second-order time-reversal symmetric TSC (class DIII) with a helical 1D mode. The second stage involves applying a TRS-breaking perturbation leading to a third-order TSC with broken TRS (class D) with a pair of Majorana zero modes at two antipodal points.}
\label{3rdOrder}
\end{figure}

We can now build a third-order TSC in 3D with corner modes by applying a TRS-breaking perturbation to the time-reversal invariant second-order TSC constructed above. The only possible PHS-preserving and TRS-breaking mass term that can be added to gap out the 1D helical mode on the surface is given by $\tilde m_\br \gamma_y$ which satisfies $\tilde m_{-\br} = -\tilde m_\br$ due to inversion. This means that the 1D helical surface mode can be gapped out except at two antipodal points where both surface mass terms $m_\br$ and $\tilde m_\br$ have to vanish. These two points resemble magnetic vortices (the two mass terms are similar to the real and imaginary part of a gap function) and they will host a Majorana mode each, similar to the vortices in a $p_x \pm i p_y$ superconductor. The construction is schematically illustrated in Fig.~\ref{3rdOrder}.

\subsection{Higher-order two-dimensional topological superconductors}
\label{2DTSC}

In analogy to the previous two sections, we can build a second-order 2D TSC in class D (superconductor with broken TRS) by applying a TRS-breaking perturbation to a 2D time-reversal-symmetric TSC (class DIII). Such a 2D TSC can be implemented as a $p_x$ (or $p_y$) wave superconductor, as a superposition of a $(p_x+ip_y)$-superconductor and its time-reversed copy \cite{Schnyder09}, or by coupling the surface of a 3D TI to a SC as shown by Fu and Kane \cite{Fu08}. Although these implementations all correspond to the AZ class DIII, the first two differ from the last one in the commutation relations between inversion and particle-hole symmetries. Whereas a $p$-wave superconductor is an odd parity superconductor with inversion anti-commuting (commuting) with PHS for $\I^2 = +1$ ($\I^2 = -1$), a Fu-Kane superconductor is constructed by proximity coupling to an $s$-wave superconductor where inversion commutes (anti-commutes) with PHS for $\I^2 = +1$ ($\I^2 = -1$). Furthermore, both implementations $\I^2 = + 1$ and $\I^2 = -1$ are possible in 2D. The former corresponds to 3D spatial inversion restricted to the 2D plane (which also flips the normal to the plane), which is relevant for example in a layered material with an inversion center in the middle, while the latter represents strictly 2D inversion which is equivalent to two-fold rotation about the axis perpendicular to the plane. According to Table~\ref{Classification}, a second-order TSC in class D is only possible when inversion anti-commutes (commutes) with PHS for $\I^2 = +1$ ($\I^2 = -1$), which means that it is only possible starting from a $p$-wave superconductor either in a layered (quasi-2D) setting with 3D inversion or in a strictly 2D system with $C_2$ symmetry. The two cases are related by the replacement $\I_{\rm 2D} = i \I_{\rm 3D}$, and in the following we will only focus on the layered (quasi-2D) setting.

We start with the Hamiltonian describing a superposition of $p_x+ip_y$ and $p_x - i p_y$ superconductors. Using the gap function $\Delta(\bk) = -\sigma_z k_x +i \sigma_0 k_y$, we can write this Hamiltonian as
\beq
\label{HDIII}
\H = -k_x \sigma_z \tau_x - k_y \tau_y + \lambda \tau_z.
\eeq 
Here, we assume the $\bk$-dependence of the single particle dispersion is weak close to 0 and replace it by a constant $\lambda$. This Hamiltonian has PHS given by $\P = \sigma_0 \tau_x \K$, TRS given by $\T = \sigma_y \tau_0 \K$, and inversion symmetry given by $\I = \sigma_0 \tau_z$. 

The combination of TRS and PHS prohibits any mass term. The only allowed TRS-breaking mass term is $m_\br \sigma_y \tau_x$, which satisfies $m_{-\br} = -m_\br$ due to inversion. As a result, it has to vanish at least twice at two inversion-related points on any given compact inversion-symmetric edge. We can see this more explicitly by writing the edge theory following the same procedures as in Secs.~\ref{3DTI} and \ref{3DTSC} leading to
\beq
h = - k_S \sigma_z,
\eeq
where $k_S$ is the momentum parallel to the edge. The edge implementation of TRS, PHS and inversion is given respectively by $\T_S = \sigma_y \K$, $\P_S = -\bn_\br \cdot \bsigma \sigma_y \K$, and $\I_S = \sigma_0$, where $\bn_\br$ is the normal to the edge in the 2D plane. The only possible TRS-breaking mass term consistent with PHS is $m_\br \bn_\br \cdot \bsigma$ which we can identify as a Zeeman term for an in-plane field. The field will fail to gap out the Hamiltonian at the two points where it is tangent to the edge, which will host a single Majorana mode each. Using this phase, we can also build a second-order 2D TSC in classes BDI or DIII by adding it to a time-reversed copy of itself with spinful or spinless TRS, respectively.

%
%
%
%

\section{Layer construction}
\label{Layer}
To establish the conditions in Table \ref{Classification}, we make use of the layer construction introduced in Refs.~\onlinecite{Isobe15,Hermele17, Huang17} to study topological phases protected by point group symmetries. Before presenting the general argument, we will first illustrate it using the example of a second-order 3D TI with unbroken TRS (class AII). Being a symmetry protected topological phase means that it becomes trivial in the absence of inversion symmetry. Let us now consider an inversion symmetric plane (i.e. one that contains the inversion center), which divides the 3D space into two regions mapped to each other under inversion. Each region separately does not have inversion symmetry and can, therefore, be adiabatically trivialized, implying that the non-triviality of the phase is only encoded in the 2D plane. To get 1D helical modes, this 2D plane needs to be a strong TI, which in this case is just a quantum-spin Hall (QSH) system. The same argument can be applied to construct a third-order TI/TSC with 0D states, which is not possible in class AII but, for instance, in the superconducting class D. In this case, we have to consider a 2D layer which hosts an inversion-protected second-order TSC, rather than a strong TSC. The reduction can then be performed further by splitting the plane into two halves using a line through the inversion center and considering a 1D strong TSC, e.g. an SSH chain, on this line to obtain a 3D third-order TSC. This argument can be generalized to relate any inversion-protected $k$-th order TI/TSC in $d$ dimensions to a $(k-1)$-th order TI/TSC in $(d-1)$ dimensions. The procedure can then be repeated to eventually reduce it to a $(d-k+1)$-dimensional strong TI/TSC.

The construction of the previous paragraph can also be reversed. Starting with an inversion-symmetric 2D QSH layer, we can build a second-order 3D TI in class AII by stacking 2D layers. To ensure that inversion symmetry is preserved, each step of the stacking consists of adjoining the system with a 2D layer (which can either be QSH or trivial insulating layers) and its copy under inversion. The procedure is shown schematically in Fig.~\ref{Layer}, where we assumed for simplicity that all layers are QSHs which are arranged such that they are equally spaced along the stacking direction. The resulting system is a 3D weak TI, whose gapless surface states are protected by translation. Breaking translation while preserving inversion will gap out the surface states except for a single 1D helical edge mode living on an inversion-symmetric curve.

\begin{figure}
\center
\includegraphics[width=0.95\columnwidth]{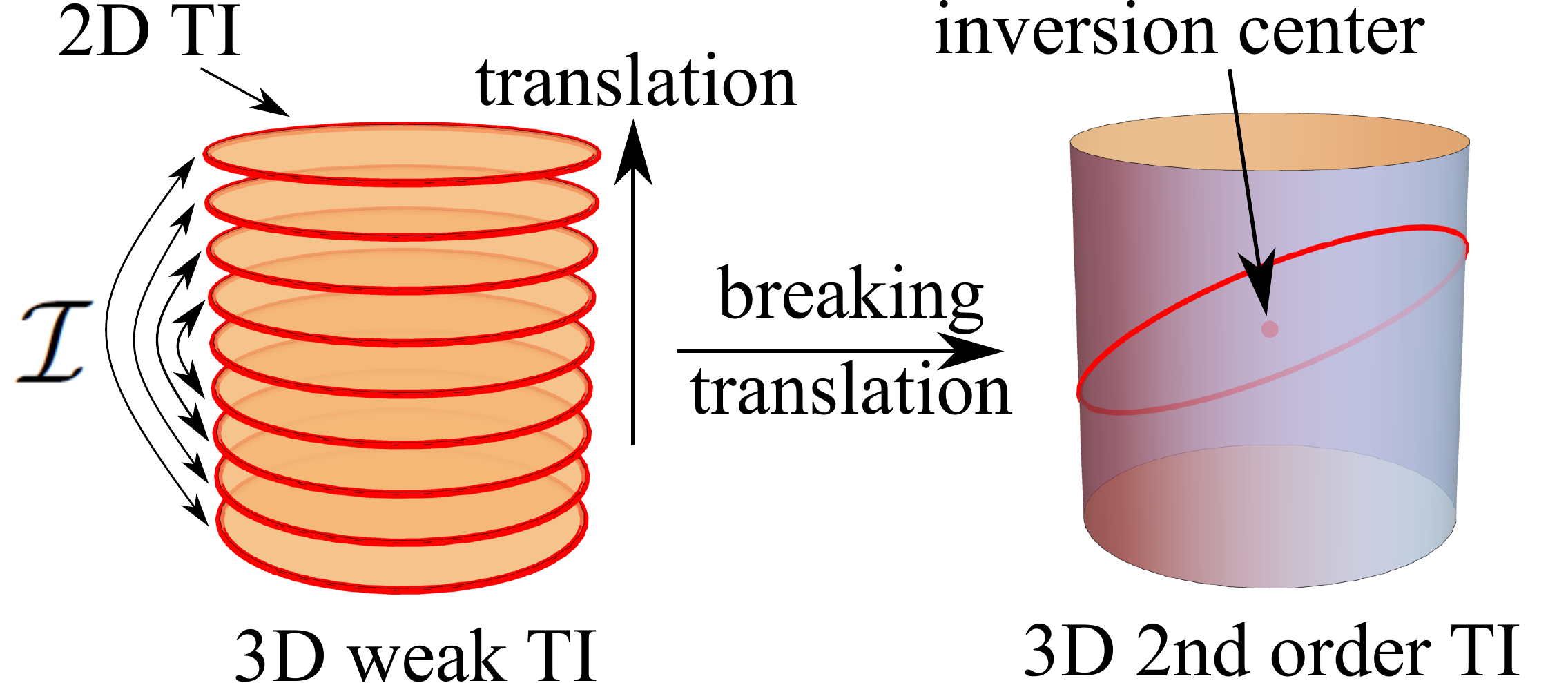}
\caption{Schematic illustration of the layer construction. Here, we can construct a second-order TI in 3D starting with a 2D strong (first order) TI and adjoining it with 2D layers and their images under inversion. We choose here to do it such that the resulting system has translational symmetry along the stacking direction, thus realizing a 3D weak TI. Breaking the translational symmetry while keeping inversion leads to a second-order TI with a single 1D mode which lives on some inversion-symmetric curve on the surface.}
\label{LayerF}
\end{figure}

In general, the construction of an inversion-protected $k$-th order TI/TSC in $d$ dimensions starts with a $(d-k+1)$-dimensional strong TI/TSC as a building block and successively adjoins it with $(d-k+1)$-dimensional blocks and their inverted copies. This suggests that the problem of classifying inversion-{\it protected} higher-order TIs/TSCs reduces to the problem of classifying inversion-{\it symmetric} strong (first-order) TIs/TSCs, i.e. those which are compatible with inversion symmetry. It should be noted, however, that not every lower-dimensional TI/TSC compatible with the symmetry would lead to a unique higher-dimensional higher-order TI/TSC. The reason is that two lower-dimensional strong phases related by the adjoining operation, which adds to the system a ``layer'' and its copy under inversion, lead to the same higher-dimensional phase and should thus be identified \cite{Hermele17}. This means that any $k$-th order TI/TSC which can be decomposed into a sum of two systems related by inversion is trivial from the point of view of higher dimensional topology and cannot be used to build any $k'$-th order phase ($k' > k$) in a higher dimension.

The resulting classification criterion for higher-order TI/TSC is the following: \\
{\it Inversion-protected $k$-th order TIs/TSCs in $d$ dimensions are in one-to-one correspondence with non-separable inversion-protected $(k-1)$-th order TIs/TSCs in $d-1$ dimensions. A non-separable TI/TSC is one which cannot be written as a sum of a TI/TSC (with the same local symmetries) and its copy under inversion.} 

The aforementioned criterion leads immediately to the anticipated $\Z_2$ classification since the sum of two inversion-symmetric strong TIs/TSCs can always be thought of as a sum of a strong TI/TSC and its copy under inversion, which is equivalent to a trivial system under the adjoining operation. We can also easily understand some of the constraints in Table \ref{Classification}. For instance, inversion symmetry is always required to anti-commute with the chiral symmetry in order for higher-order TIs to be possible in class AIII. This can be easily seen from the fact that strong TIs in class AIII are classified with an integer winding number in odd dimensions $d$ given by $\sim \int \tr \S (\H^{-1} d \H)^{2n+1}$ \cite{Ryu10, Shiozaki14},
which is even (odd) under inversion if it anti-commutes (commutes) with the chiral symmetry $\S$. Therefore, a non-vanishing winding number cannot be consistent with an inversion symmetry which commutes with $\S$.

We now describe how this criterion can be used to establish the classification of Table \ref{Classification}. This is done by first considering all non-separable inversion-symmetric strong TIs/TSCs. We then construct higher-order TIs/TSCs order by order using a dimensional raising map similar to the one used in Refs.~\onlinecite{Teo10, Shiozaki14} as we will show below. The main subtlety in this construction is that we need to ensure that the Hamiltonian is non-separable at every stage of the procedure\footnote{I am grateful to L.~Trifunovic and P.~Brouwer for pointing this out to me}. This problem can be simplified by restricting ourselves to analyzing Dirac Hamiltonian (DH) representatives of each phase. Such restriction is justified by the observation that the Hamiltonian of any higher-order TI/TSC can be deformed to a DH provided that we allow for the addition or removal of bands which do not contribute to the surface states at all and are thus ``trivial'' from the surface states point of view. This means that, unlike the classification of strong TIs/TSCs \cite{Ryu10} and TCIs \cite{Shiozaki14}, we identify phases with the same surface states {\it even if} they are not adiabatically deformable.

Let us now first review the DH description for strong TIs/TSCs. The ten AZ symmetry classes are divided into two complex and eight real classes. The former includes classes which do not posses any antiunitary symmetry relating the first quantized Hamiltonian to its complex conjugate. These are unitary (A) and chiral unitary (AIII) classes which are denoted by the complex $s_c$ parameter $s_c = 0$ and $s_c = 1$ respectively (cf.~Table \ref{Classification}). Real classes include the remaining eight classes which are labeled by the parameter $s_r = 0,\dots, 7$ as shown in Table \ref{Classification}. A TI/TSC with a $\Z$ classification is possible in the complex classes whenever $s_c - d = 0 \mod 2$. The $\Z$ invariant is a Chern number in even dimensions and a winding number in odd dimensions. Examples include the SSH chain in 1D \cite{Su79} ($s_c = 1$) and the quantum Hall effect in 2D \cite{Klitzing80, Thouless82} ($s_c = 0$). For real classes, there are four series of TIs/TSCs given by $s_r - d = 0,1,2,4 \mod 8$. The series $s_r - d=0 \mod 8$ has a $\Z$ invariant given by a Chern number in even dimensions and a winding number in odd dimensions. It includes topological $p_x \pm i p_y$ TSCs in 2D \cite{Read00, Schnyder09} ($s_r = 2$) and class DIII time-reversal-symmetric TSCs in 3D \cite{Volovik03, Schnyder09} ($s_r = 3$). The first and second descendant series are given respectively by $s_r - d = 1, 2 \mod 8$ and they have $\Z_2$ classification. An example of the former is the 3D time-reversal-symmetric TI \cite{Fu07, Moore07} ($s_r = 4$) and of the latter is the quantum-spin Hall effect in 2D \cite{Kane05a, Kane05b, Bernevig06}($s_r = 4$). Finally, there is the $s_r - d = 4 \mod 8$ series which has a $2\Z$ classification, meaning that the Chern number (for even dimensions) or the winding number (for odd dimensions) is always even \cite{Kitaev09, Schnyder09, Ryu10}.

The DH for a strong (first-order) TI/TSC can be written in all cases as
\beq
\label{HD1}
\H = \sum_{i=1}^d \Gamma_i \sin k_i - \M (d - 1 - \sum_{i=1}^d \cos k_d).
\eeq
Here, $\Gamma_i$ denotes a set of Dirac matrices satisfying the Clifford algebra $\{\Gamma_i, \Gamma_j\} = 2\delta_{ij}$ with $i,j = 1, \dots, 2n+1$. Only $n$ of these $\Gamma$ matrices are imaginary and we choose these to be the even ones. The Dirac mass $\M$ always satisfies $\M^2 = 1$ and $\{\M, \Gamma_i\} = 0$ for $i = 1,\dots,d$.

The possible choices of the mass term $\M$ for different symmetry classes in different dimensions are given in Table \ref{DiracTable}, where strong TIs/TSCs correspond to $k=1$ (first-order). The anti-unitary operator $\A$ is defined for $k=1$ as
\beq
\A^{(1)} = \prod_{\substack{l=1\\ l \text{ odd}}}^{d} \Gamma_l \, \K,
\eeq
and it may commute (anti-commute) with the DH (\ref{HD1}) depending on the choice of the mass term $\M$ and the dimension, representing TRS (PHS). The 7th column in Table \ref{DiracTable} provides all possible implementations of the inversion operator $\I$ such that $\I^2 = 1$, $[\I, \M] = 0$, and $\{\I, \Gamma_i\} = 0$ for $i = 1,\dots, d$. The next three columns indicate whether it commutes ($+$) or anti-commutes ($-$) with $\T$, $\P$, and $\S$.

A DH for a $k$-th order TI/TSC in $d$ can be constructed from the DH for a $(k-1)$-th order TI/TSC in $d-1$ dimensions using the recurrence relation
\begin{gather}
\H^{(k)} = \H^{(k-1)} \otimes \tau_z - \M^{(k)} (1 - \cos k_{d}) + \sin k_{d} \tau_x, \nonumber \\
\M^{(k)} = \M^{(k-1)}\otimes \tau_z,
\label{Hk}
\end{gather}
with $\T$, $\P$, $\S$, and $\I$ symmetries given by
\begin{gather}
\T^{(k)} = \T^{(k-1)} \otimes \tau_z, \quad \P^{(k)} = \P^{(k-1)} \otimes \tau_0, \nonumber \\
\S^{(k)} = \S^{(k-1)} \otimes \tau_z, \quad \I^{(k)} = \I^{(k-1)} \otimes \tau_z.
\end{gather}
Compared to $\H^{(k-1)}$, the Hamiltonian $\H^{(k)}$ admits an extra symmetry-allowed mass term $\sim \tau_y$, which is odd under inversion. Since no symmetry-allowed mass can be added to $\H^{(1)}$ (which describes a strong TI/TSC), we can conclude by induction that $\H^{(k)}$ has $k-1$ symmetry-allowed mass terms which are odd under inversion. These mass terms will gap-out the surface leaving a gapless $(d-k)$-dimensional region, thus implementing a $k$-th order TI/TSC. This will be explained in more detail in the next section. 

To establish the classification in Table \ref{Classification}, we need to check whether the Hamiltonian $\H^{(k)}$ is separable i.e. whether it can be written as the sum of two Hamiltonians (with the same local symmetries) related to each other by inversion. Separability of a Hamiltonian $\H$ is equivalent to the existence of a unitary operator $U$ with $U^2 = 1$ satisfying
\beq
\label{Uk}
[U,\H] = [U,\T] = [U,\P] = [U,\S] = \{U,\I \} = 0.
\eeq
If such operator exists, we can decompose the Hamiltonian as 
\beq
\label{Hpm}
\H = \H_{+} + \H_{-}, \qquad \H_{\pm} = \frac{1}{2} (1 \pm U) \H.
\eeq
$\H_{\pm}$ satisfy the same local symmetries ($\T$, $\P$ and $\S$) as $\H$. In addition, they map to each other under inversion since
\beq
\label{IHpm}
\I^{-1} \H_{\pm} \I = \H_{\mp},
\eeq
which implies that $\H$ is separable. Conversely, if the Hamiltonian can be decomposed as in (\ref{Hpm}) with $\H_{\pm}$ satisfying (\ref{IHpm}), we can choose a basis where $\I$ is represented as $\sigma_x$ so that $\H$ has the form
\beq
\H = \left(\begin{array}{cc} \H_{+} & 0 \\ 0 & \H_{-} \end{array} \right).
\eeq
In this basis, $\T$, $\P$, and $\S$ will be diagonal since they preserve $\H_+$ and $\H_-$ separately. By defining $U$ as $\sigma_z$, we see that it will satisfy the commutation relations (\ref{Uk}). We notice that the existence an operator $U$ satisfying (\ref{Uk}) for the $k$-th order Hamiltonian $\H^{(k)}$ implies that $\H^{(k+1)}$ constructed according to (\ref{Hk}) does not have any surface states. This is consistent with our criterion which implies that the existence of a {\it non-separable} $k$-th order TI/TSC in $d$ dimensions is a necessary (and sufficient) condition for the existence of a $(k+1)$-th order TI/TSC in $d+1$ dimensions, which will be impossible in this case. One direct way to see this is by noticing that the existence of the unitary operator $U$ for some $k$ implies the existence of a mass term $\M^{(k+1)} = U \otimes \tau_y$ satisfying
\begin{gather}
\{\M^{(k+1)},\H^{(k+1)}\} = [\M^{(k+1)},\I] = 0, \nonumber \\
[\M^{(k+1)},\T] = \{\M^{(k+1)},\P\} = \{\M^{(k+1)},\S\} = 0,
\end{gather}
which can be used to completely gap-out the surface. Such mass term can be used to construct similar mass terms $\M^{(k')}$ for $k' > k$ using the relation $\M^{(k'+1)} = \M^{(k')} \otimes \tau_z$, thus ruling out the existence of any $k'$-th order TI/TSC for $k' > k$.

The results of the analysis outlined above are summarized in Table \ref{DiracTable}. The DH Hamiltonian for a $k$-th order TI/TSC in $d$-dimensions is given by the same expression (\ref{HD1}) with the Dirac mass $\M$ and the symmetries $\T$, $\P$, $\S$, and $\I$ provided in columns 3 to 7. The operator $\A$ for $k>1$ is given by
 \beq
 \label{A}
\A^{(k)} = \prod_{\substack{l=1\\ l \text{ odd}}}^{d} \Gamma_l \prod_{\substack{m=d+1\\ m \text{ even}}}^{d+k-1} \Gamma_m \, \K.
\eeq
The last column indicates the separability matrix $U$ which can be used to decompose the Hamiltonian as in (\ref{Hpm}) (0 indicates that no such $U$ exists for any order $k$) and the column before indicates the minimum value of the order $k=k_m$ for which this is possible. Finite $k_m$ means that a $k$-the order TI/TSC is impossible whenever $k > k_m$. We notice that whenever the inversion operator is the same as the mass term $\I = \M$, $k_m$ is infinite and the Hamiltonian for the $k$-th order TI/TSC is non-separable for any $k$. For any other choice of the inversion operator, $k_m$ is finite and the Hamiltonian is only non-separable for $k<2$ or $k<3$ depending on the symmetry class and dimension (cf.~Table \ref{DiracTable}). 

\section{Surface theory for higher-order topological insulators and superconductors in arbitrary dimension}
\label{Dirac}

In this section, we show that the minimal Dirac model for $k$-th order TIs/TSCs provided in the previous section does indeed posses $(d-k)$-dimensional surface states. The main observation is that such a model admits $k-1$ extra mass terms, which are odd under inversion. The surface states can be analyzed by expanding the Hamiltonian (\ref{HD1}) around $\bk = 0$ and including all possible symmetry-allowed mass terms. For a $k$-th order TI/TSC in $d$ spatial dimensions with $k \leq k_m$, this leads to
\begin{align}
\label{HDk}
\H &= -\sum_{i=1}^d i \Gamma_i \partial_i + \sum_{j=1}^{k-1} m_{j,\br} \Gamma_{d+j} + \lambda \M \nonumber \\
&= -i \bgamma \cdot \bnabla + \bm_\br \cdot \balpha + \lambda \M,
\end{align}
where we introduced the vectors $\bgamma$, $\bm$ and $\balpha$ defined as
\begin{gather}
\bgamma = (\Gamma_1, \dots, \Gamma_d), \quad \balpha = (\Gamma_{d+1}, \dots, \Gamma_{d+k-1}), \nonumber \\
\bm_\br = (m_{1,\br},\dots,m_{k-1,\br}). 
\end{gather}
The explicit form of the Dirac mass and the symmetries $\T$, $\P$, $\S$, and $\I$ are provided in Table \ref{DiracTable}.
The inversion operator anti-commutes with all $\Gamma_l$ for $l = 1,\dots, d + k-1$ and commutes with $\M$. As a result, it enforces the condition $m_{j,-\br} = - m_{j,\br}$.

In order to see the type of surface states corresponding to the DH (\ref{HDk}), we follow the same procedure of Sec.~\ref{Realizations} and take $\lambda$ to change from $1$ inside the sample to $-1$ outside it across some inversion-symmetric surface. We then introduce the projection operators
\beq
P_{\pm} = \frac{1}{2} (1 \mp i \, \bn_\br \cdot \bgamma \, \M), \quad P_\pm^2 = P_\pm, \quad P_+ P_- = 0,
\eeq
where $\bn_\br$ is the normal to the surface at point $\br$. Using the projector $P_+$, the surface Hamiltonian can be obtained as
\beq
h =  \tilde \bgamma \cdot \bk_S + \bm_\br \cdot \tilde \balpha, \quad \tilde \bgamma = P_+ \bgamma P_+, \quad \tilde \balpha = P_+ \balpha P_+,
\eeq
where $\bk_S$ is the momentum tangent to the surface. The projector $P_+$ ensures that the momentum perpendicular to the surface drops out since 
\beq
\bn_\br \cdot \tilde \bgamma =  P_+ \bn_\br \cdot \bgamma P_+ = P_+ P_- \bn_\br \cdot \bgamma = 0,
\eeq
which follows from the relation $P_+ \bn_\br \cdot \bgamma = \bn_\br \cdot \bgamma P_-$ (since $\M$ anticommutes with $\bn_\br \cdot \bgamma$). 

We now consider an orthonormal basis $\{\be_i\}$ in the $(d-1)$-dimensional plane tangent to the surface at a given point $\br$, i.e. $\be_i \cdot \be_j = \delta_{ij}$, and define $\tilde \gamma_i = \be_i \cdot \tilde \bgamma$. It is easy to see that 
\beq
\{\tilde \gamma_i, \tilde \gamma_j\} = 2\delta_{ij} P_+, \quad \{\tilde  \gamma_i, \tilde \alpha_l\} = 0, \quad \{ \tilde \alpha_l, \tilde \alpha_m\} = 2\delta_{lm} P_+,
\eeq
for $i, j = 1,\dots,d-1$ and $l,m = 1,\dots,k-1$. This means that $\tilde \gamma_i$ and $\tilde \alpha_l$ form a Clifford algebra once projected to the non-zero block of the projector $P_+$ and the surface Hamiltonian has the spectrum
\beq
\epsilon_{\br,\bk_S} = \pm \sqrt{\bk_S^2 + \bm_{\br}^2},
\eeq
which is gapless whenever all the masses $m_{j,\br}$ vanish. Each of them is odd under inversion $m_{j,-\br} = -m_{j,\br}$ and thus vanishes on a $(d-2)$-dimensional region on the surface. It follows that they simultaneously vanish at a $(d-k)$-dimensional region on the surface, thus realizing a $k$-th order TI/TSC.

\section{Discussion}
\label{Discussion}

We now make a few closing remarks regarding the stability of the phases considered in this work, its relation to other works, and the generalization of the analysis considered here for other spatial symmetries.

First, we note that, similar to other gapped topological phases of matter, the phases considered here are protected by the bulk gap. This means that symmetry-preserving perturbations that are small enough compared to the bulk gap will not be able to destroy the surface states. They can, however, move the surface states around as explained in the main text. We note also that the surface states are stable against inversion-breaking perturbations, e.g. disorder, provided they are small compared to the maximal value of the surface gap. The reason for this is that the local stability of the surface states relies only on the local symmetries $\T$, $\P$, and $\S$. Therefore, the only way to remove them is by deforming them all the way to a point, for example by bringing two corner states together or deforming a 1D surface state to a point. 
In summary, bulk gap provides protection against inversion-preserving perturbations, while surface gap provides protection against inversion-breaking perturbations. This conclusion is consistent with the analysis of Ref.~\onlinecite{Langbehn17}, which showed that the surface states in a mirror-protected second-order TIs/TSCs are stable against mirror-symmetry-breaking perturbations. Note, however, that perturbations that break the local symmetries $\T$, $\P$, and $\S$ can generally destabilize the surface states even if they are very small.

Second, we reiterate here that we do not provide a classification of all TCIs protected by inversion (such classification was obtained using K-theory in Refs.~\onlinecite{Shiozaki14, Lu14}). Instead, we distinguish TCIs by their pattern of surface states. This means that two distinct K-theory phases with the same pattern of surface states are considered equivalent even if they are not adiabatically deformable to each other. This could happen if they differ by the addition of a non-trivial K-theory phase with completely gapped surface. The existence of such phases can be seen by considering the Dirac representatives provided in Ref.~\onlinecite{Lu14}, whose Hamiltonians have a very similar form to Eq.~\ref{HDk}, and noting that the surface can be gapped completely whenever the number of possible symmetry-allowed mass terms at the surface equals or exceeds the spatial dimension. Examples of such phases include obstructed atomic limits (or frozen polarization phases) \cite{Bradlyn17, Po17} in the absence of chiral or particle-hole symmetry to stabilize potential edge or corner modes, e.g. SSH chain with inversion. Whether all K-theory TCIs without surface states correspond to atomic insulators is, to our knowledge, an open question. 

It follows from the previous discussion that the set of phases obtained here is equivalent to the K-theory phases modulo those without surface states. We stress that such phases exhaust all possible surface states protected by inversion symmetry in the 10 AZ symmetry classes in any dimension. Unlike other spatial symmetries which leave some surface planes invariant, e.g. mirror symmetry, inversion does not leave any plane on the surface invariant. Therefore, inversion-protected surface states can only be observed by considering a sample with compact geometry and open boundary conditions in all directions, where the surface is considered as a whole.

It is worth noting that the layer construction employed in this work was originally employed to study topological phases protected by spatial symmetries in the presence of interactions \cite{Hermele17, Huang17}. Although the effect of interactions is beyond the scope of this work, we can make the following remarks. Generally, the addition of interactions can change the non-interacting classification in two ways: (i) it can destabilize some of the non-interacting phases leading to a completely gapped surface, thereby reducing the non-interacting classification \cite{Fidkowski10, Fidkowski11,Ryu12, Qi13, Yao13, Wang14, Gu14}, or (ii) it can introduce new phases that do not have any non-interacting counterparts \cite{Wang14b, Lapa16}. In conventional TIs/TSCs, it is known that the classification reduction only happens for $\Z$ phases with chiral symmetry, which are reduced to $\Z_n$ (for some even integer $n$) when symmetry-preserving interactions are added. Such interactions introduce coupling between the surface modes which gaps them out \cite{Morimoto15, Queiroz16}. This suggests that the phases considered here, which are all $\Z_2$ phases hosting a single surface mode, are stable against interactions. We can see this more directly by noting that the generator element of the $\Z$ or $\Z_2$ factor (in the conventional TI/TSC classification) used to build inversion-protected higher-order TIs/TSCs (cf.~Sec.~\ref{Layer}) is generally stable to the addition of interactions. The possibility of interaction-induced phases is more difficult to investigate, but we expect the main conclusions of this work to hold. Mainly, that such interacting inversion-protected higher-order phases are built by embedding inversion-symmetric interacting phases (with no non-interacting counterparts) in higher-dimensions using the layer construction and that the resulting classification is always $\Z_2$.

Finally, we note that the method we used here, which combines a Dirac analysis with the layer construction used in Refs.~\onlinecite{Hermele17, Huang17,SongFang17}, can be readily generalized to any spatial symmetry and it does not require the knowledge of the full K-theory classification of the corresponding TCIs, which is only known for order-two symmetry operations \cite{Shiozaki14}. The method employed here was recently used to classify all possible surface states in TCIs with strong spin-orbit coupling (class AII) protected by any crystalline symmetry in the 230 space groups \cite{Khalaf17}. Although the extension of our results to include other spatial symmetries, e.g. rotations, is straightforward, the analysis is more complicated since special care is needed when considering points, lines, or planes left invariant by the symmetry. In addition, the existence of surface states may, in many cases, require unphysical commutation or anticommutation relations between time-reversal or particle-hole symmetries and spatial symmetries. A careful investigation would then be required to separate the physically more relevant cases as we have done in this work. Furthermore, higher-order TIs/TSCs protected by symmetries with invariant planes (which includes all other point group symmetries apart from roto-inversion) can usually be understood in a more conventional way by studying possible surface states on these planes. Therefore, we choose to leave this question to future works.\\

{\it Note added.} After the completion of this work, the author became aware of two related studies: one considers second-order topological insulators and superconductors protected by order-two symmetries \cite{Geier18} and the other discusses realizations of the two-dimensional second-order topological superconductor considered in this work \cite{Zhu18}.

\begin{acknowledgements}
The author would like to acknowledge stimulating discussions with P.~W.~Brouwer, M.~Geier, A.~P.~Schnyder, H.~C.~Po, and A.~Vishwanath. I am very grateful to L.~Trifunovic and P.~W.~Brouwer for pointing out an error in an earlier version of this manuscript which led to some errors in the classification table and for informing me of their related upcoming work. I am also grateful to E.~Locane for proofreading the manuscript.
\end{acknowledgements}




\bibliography{refs}

\begin{center}
\begin{table*}
\caption{{\bf Data for the Dirac Hamiltonian description for higher-order TIs/TSCs}. The first column gives the AZ label for the five series of higher-order TIs/TSCs $s_c - \delta = 0 \mod 2$ (complex) and $s_r - \delta = 0,1,2,4 \mod 8$ (real), with $\delta$ given by $\delta = d - k + 1$ as in Table \ref{Classification} and $\td$ given by $\td=d+k$. The third column gives the Dirac mass for the Hamiltonian (\ref{HD1}) consistent with the given symmetry class and dimension. The next three columns provide the explicit implementation for time-reversal, particle-hole and chiral symmetries, in terms of the anti-unitary operator $\A$ defined in (\ref{A}). The 7th column gives all possible implementations of the inversion symmetry $\I$ such that it satisfies $\I^2 = 1$, $[\I,\M] = 0$, and $\{\I, \Gamma_i\} = 0$ for $i = 1,\dots,\td-1$. The last three column give the commutation/anti-commutation sign of inversion with time-reversal, particle-hole and chiral symmetries defined as $\sigma_\T = \I \T^{-1} \I \T$, $\sigma_\P = \I \P^{-1} \I \P$, and $\sigma_\S = \I \S^{-1} \I \S$, respectively. The last two columns indicate the maximal order $k_m$ such that a $k$-th order phase is possible for all $k \leq k_m$ and the unitary separation matrix $U$ defined in (\ref{Uk}) for $k=k_m$.}
\bgroup
\setlength\extrarowheight{0.25em}
\setlength{\tabcolsep}{0.1em}
\begin{tabular}{c|c|c|c|c|c|c|c|c|c|c|c}
\hline \hline
$\! s_c \! - \!\delta \! \! \! \mod \! 2$ & $\delta \!\! \! \mod 2$ & $\M$ & $\T$ & $\P$ & $\S$ & $\I$ & $\sigma_\T$ & $\sigma_\P$ & $\sigma_\S$ & $k_m$ & $U$\\
\hline
\multirow{2}{*}{0} & 0 & $\Gamma_{\td}$ & 0 & 0 & 0 & $\Gamma_{\td}$ & 0 & 0 & 0 & $\infty$ & 0\\
\cline{2-12}
 & 1 & $\Gamma_{\td+1}$ & 0 & 0 & $\Gamma_{\td}$ & $\Gamma_{\td+1}$ & 0 & 0 & $-$ & $\infty$ & 0\\
\hline \hline
$\! s_r \! - \! \delta \! \! \! \mod \! 8$ & $\delta \!\! \! \mod\! 4$ & $\M$ & $\T$ & $\P$ & $\S$ & $\I$ & $\sigma_\T$ & $\sigma_\P$ & $\sigma_\S$ & $k_m$ & $U$\\
\hline
\multirow{4}{*}{0} & 0 & $\Gamma_{\td}$ & $\A$ & 0 & 0 & $\Gamma_{\td}$ & $+$ & 0 & 0 & $\infty$ & 0\\
\cline{2-12}
& 1 & $\Gamma_{\td+1}$ & $\Gamma_{\td} \A$ & $\A$ & $\Gamma_{\td}$ & $\Gamma_{\td+1}$ & $+$ & $-$ & $-$ & $\infty$ & 0\\
\cline{2-12}
  & 2 & $\Gamma_{\td}$ & 0 & $\A$ & 0 & $\Gamma_{\td}$ & 0 & $-$ & 0 & $\infty$ & 0\\
  \cline{2-12}
   & 3 & $\Gamma_{\td+1}$ & $\A$ & $\Gamma_{\td} \A$ & $\Gamma_{\td}$ & $\Gamma_{\td+1}$ & $+$ & $-$ & $-$ & $\infty$ & 0\\
 \hline
 \multirow{6}{*}{1} & \multirow{2}{*}{0} & \multirow{2}{*}{$\Gamma_{\td+2}$} & \multirow{2}{*}{$\A$} & \multirow{2}{*}{$\Gamma_{\td} \A$} & \multirow{2}{*}{$\Gamma_{\td}$} & $\Gamma_{\td+2}$ & $+$ & $-$ & $-$ & $\infty$ & 0\\
 \cline{7-12}
 &&&&&&$i\Gamma_{\td} \Gamma_{\td+1} \Gamma_{\td+2}$ & $+$ & $+$ & $+$ & 2 & $i \Gamma_{\td-1} \Gamma_{\td-1}$\\
 \cline{2-12}
 & 1 & $\Gamma_{\td+1}$ & 0 & $\A$ & 0 & $\Gamma_{\td+1}$ & 0 & $-$ & 0 & $\infty$ & 0\\
 \cline{2-12}
   & \multirow{2}{*}{2} & \multirow{2}{*}{$\Gamma_{\td+2}$} & \multirow{2}{*}{$\Gamma_{\td} \A$} & \multirow{2}{*}{$\A$} & \multirow{2}{*}{$\Gamma_{\td}$} & $\Gamma_{\td+2}$ & $+$ & $-$ & $-$ & $\infty$ & 0\\
   \cline{7-12}
   &&&&&&$i\Gamma_{\td} \Gamma_{\td+1} \Gamma_{\td+2}$ & $-$ & $-$ & $-$ & 2 & $i \Gamma_{\td-1} \Gamma_{\td+1}$\\
   \cline{2-12}
    & 3 & $\Gamma_{\td+1}$ & $\A$ & 0 & 0 & $\Gamma_{\td+1}$ & $+$ & $0$ & $0$ & $\infty$ & 0\\
  \hline
 \multirow{12}{*}{2} & \multirow{2}{*}{0} & \multirow{2}{*}{$\Gamma_{\td+2}$} & \multirow{2}{*}{0} & \multirow{2}{*}{$\Gamma_{\td} \A $} & \multirow{2}{*}{0} & $\Gamma_{\td+2}$ & 0 & $-$ & 0 & $\infty$ & 0\\
   \cline{7-12}
  &&&&&& $i \Gamma_{\td} \Gamma_{\td+1} \Gamma_{\td+2}$ & 0 & $+$ & 0 & 2 & $i \Gamma_{\td-1} \Gamma_{\td+1}$\\
  \cline{2-12}
 & \multirow{4}{*}{1} & \multirow{4}{*}{$\Gamma_{\td+3}$} & \multirow{4}{*}{$\Gamma_{\td+1} \A$} & \multirow{4}{*}{$\A$} & \multirow{4}{*}{$\Gamma_{\td+1}$} & $\Gamma_{\td+3}$ & $+$ & $-$ & $-$ & $\infty$ & 0\\
   \cline{7-12}
 &&&&&& $i\Gamma_{\td} \Gamma_{\td+1} \Gamma_{\td+3}$ & $-$ & $-$ & $+$ & 2 & $i \Gamma_{\td-1} \Gamma_{\td}$\\
   \cline{7-12}
 &&&&&& $i\Gamma_{\td} \Gamma_{\td+2} \Gamma_{\td+3}$ & $-$ & $+$ & $-$ & 2 & $i \Gamma_{\td-1} \Gamma_{\td}$\\
   \cline{7-12}
 &&&&&& $i\Gamma_{\td+1} \Gamma_{\td+2} \Gamma_{\td+3}$ & $-$ & $-$ & $+$ & 2 & $i \Gamma_{\td-1} \Gamma_{\td+2}$\\
 \cline{2-12}
   & \multirow{2}{*}{2} & \multirow{2}{*}{$\Gamma_{\td+2}$} & \multirow{2}{*}{$\Gamma_{\td} \A $} & \multirow{2}{*}{0} & \multirow{2}{*}{0} & $\Gamma_{\td+2}$ & $+$ & 0 & 0 & $\infty$ & 0\\
     \cline{7-12}
  &&&&&& $i \Gamma_{\td} \Gamma_{\td+1} \Gamma_{\td+2}$ & $-$ & 0 & 0 & 2 & $i \Gamma_{\td-1} \Gamma_{\td}$\\
  \cline{2-12}
    & \multirow{4}{*}{3} & \multirow{4}{*}{$\Gamma_{\td+3}$} & \multirow{4}{*}{$\A$} & \multirow{4}{*}{$\Gamma_{\td+1} \A$} & \multirow{4}{*}{$\Gamma_{\td+1}$} & $\Gamma_{\td+3}$ & $+$ & $-$ & $-$ & $\infty$ & 0\\
      \cline{7-12}
   &&&&&& $i\Gamma_{\td} \Gamma_{\td+1} \Gamma_{\td+3}$ & $+$ & $+$ & $+$ & 2 & $i \Gamma_{\td-1} \Gamma_{\td}$\\
     \cline{7-12}
   &&&&&& $i\Gamma_{\td} \Gamma_{\td+2} \Gamma_{\td+3}$ & $-$ & $+$ & $-$ & 2 & $i \Gamma_{\td-1} \Gamma_{\td}$\\
     \cline{7-12}
   &&&&&& $i\Gamma_{\td+1} \Gamma_{\td+2} \Gamma_{\td+3}$ & $+$ & $+$ & $+$ & 2 & $i \Gamma_{\td-1} \Gamma_{\td+2}$ \\
   \hline
 \multirow{16}{*}{4} & \multirow{4}{*}{0} & \multirow{4}{*}{$i \Gamma_{\td} \Gamma_{\td+1} \Gamma_{\td+2}$} & \multirow{4}{*}{$\Gamma_{\td} \Gamma_{\td+2} \A$} & \multirow{4}{*}{0} & \multirow{4}{*}{0} & $i \Gamma_{\td} \Gamma_{\td+1} \Gamma_{\td+2}$ & $+$ & 0 & 0 & $\infty$ & 0\\
      \cline{7-12}
   &&&&&& $\Gamma_{\td}$ & $-$ & 0 & 0 & 3 & $\Gamma_{\td-2} \Gamma_{\td-1} \Gamma_{\td} \Gamma_{\td+1}$\\
     \cline{7-12}
   &&&&&& $\Gamma_{\td+1}$ & $-$ & 0 & 0 & 3 & $\Gamma_{\td-2} \Gamma_{\td-1} \Gamma_{\td} \Gamma_{\td+1}$\\
     \cline{7-12}
   &&&&&& $\Gamma_{\td+2}$ & $-$ & 0 & 0 & 3 & $\Gamma_{\td-2} \Gamma_{\td-1} \Gamma_{\td} \Gamma_{\td+2}$\\
   \cline{2-12}
   & \multirow{4}{*}{1} & \multirow{4}{*}{$i \Gamma_{\td+1} \Gamma_{\td+2} \Gamma_{\td+3}$} & \multirow{4}{*}{$\Gamma_{\td} \Gamma_{\td+1} \Gamma_{\td+3} \A$} & \multirow{4}{*}{$\Gamma_{\td+1} \Gamma_{\td+3}\A$} & \multirow{4}{*}{$\Gamma_{\td}$} & $i \Gamma_{\td+1} \Gamma_{\td+2} \Gamma_{\td+3}$ & $+$ & $-$ & $-$ & $\infty$ & 0\\
      \cline{7-12}
   &&&&&& $\Gamma_{\td+1}$ & $-$ & $+$ & $-$ & 3 & $\Gamma_{\td-2} \Gamma_{\td-1} \Gamma_{\td+1} \Gamma_{\td+2}$\\
     \cline{7-12}
   &&&&&& $\Gamma_{\td+2}$ & $-$ & $+$ & $-$ & 3 & $\Gamma_{\td-2} \Gamma_{\td-1} \Gamma_{\td+1} \Gamma_{\td+2}$\\
     \cline{7-12}
   &&&&&& $\Gamma_{\td+3}$ & $-$ & $+$ & $-$ & 3 & $\Gamma_{\td-2} \Gamma_{\td-1} \Gamma_{\td+1} \Gamma_{\td+3}$\\
   \cline{2-12}
   & \multirow{4}{*}{2} & \multirow{4}{*}{$i \Gamma_{\td} \Gamma_{\td+1} \Gamma_{\td+2}$} & \multirow{4}{*}{0} & \multirow{4}{*}{$\Gamma_{\td} \Gamma_{\td+2}\A$} & \multirow{4}{*}{0} & $i \Gamma_{\td} \Gamma_{\td+1} \Gamma_{\td+2}$ & 0 & $-$ & 0 & $\infty$ & 0\\
      \cline{7-12}
   &&&&&& $\Gamma_{\td}$ & 0 & $+$ & 0& 3 & $\Gamma_{\td-2} \Gamma_{\td-1} \Gamma_{\td} \Gamma_{\td+1}$\\
     \cline{7-12}
   &&&&&& $\Gamma_{\td+1}$ & 0 & $+$ & 0& 3 & $\Gamma_{\td-2} \Gamma_{\td-1} \Gamma_{\td} \Gamma_{\td+1}$\\
     \cline{7-12}
   &&&&&& $\Gamma_{\td+2}$ & 0 & $+$ & 0 & 3 & $\Gamma_{\td-2} \Gamma_{\td-1} \Gamma_{\td} \Gamma_{\td+2}$\\
   \cline{2-12}
   & \multirow{4}{*}{3} & \multirow{4}{*}{$i \Gamma_{\td+1} \Gamma_{\td+2} \Gamma_{\td+3}$} & \multirow{4}{*}{$\Gamma_{\td+1} \Gamma_{\td+3} \A$} & \multirow{4}{*}{$\Gamma_{\td} \Gamma_{\td+1} \Gamma_{\td+3}\A$} & \multirow{4}{*}{$\Gamma_{\td}$} & $i \Gamma_{\td+1} \Gamma_{\td+2} \Gamma_{\td+3}$ & $+$ & $-$ & $-$ & $\infty$ & 0\\
      \cline{7-12}
   &&&&&& $\Gamma_{\td+1}$ & $-$ & $+$ & $-$ & 3 & $\Gamma_{\td-2} \Gamma_{\td-1} \Gamma_{\td+1} \Gamma_{\td+2}$\\
     \cline{7-12}
   &&&&&& $\Gamma_{\td+2}$ & $-$ & $+$ & $-$ & 3 & $\Gamma_{\td-2} \Gamma_{\td-1} \Gamma_{\td+1} \Gamma_{\td+2}$\\
     \cline{7-12}
   &&&&&& $\Gamma_{\td+3}$ & $-$ & $+$ & $-$ & 3 & $\Gamma_{\td-2} \Gamma_{\td-1} \Gamma_{\td+1} \Gamma_{\td+3}$\\
 \hline \hline
\end{tabular}
\egroup
\label{DiracTable}
\end{table*}
\end{center}
\end{document}